\documentclass{emulateapj}

\shorttitle{C$_{60}$ in PDRs}
\shortauthors{Castellanos et al.}

\newcommand{\dif}[1]{\mathrm{d}#1}

\begin{document}

\title{C$_{60}$ in Photodissociation Regions}

\author{Pablo Castellanos$^{1}$, Olivier Bern\'e$^{2,3}$, Yaron
  Sheffer$^{4}$, Mark G.\ Wolfire$^{4}$ and Alexander G.G.M.\
  Tielens$^{1}$} 

\affil{$^{1}$Leiden Observatory, Leiden University, P.O. Box 9513,
  2300 RA Leiden, The Netherlands}

\affil{$^{2}$Universit\'e de Toulouse, UPS-OMP, IRAP, F-31400 Toulouse,
  France}

\affil{$^{3}$CNRS, IRAP, 9 Av.\ Colonel Roche, BP 44346, 31028 Toulouse
  Cedex 4, France} 

\affil{$^{4}$Department of Astronomy, University of Maryland, College Park,
  MD 20742, USA} 

\email{pablo@strw.leidenuniv.nl}

\begin{abstract}
  Recent studies have confirmed the presence of buckminsterfullerene
  (C$_{60}$) in different interstellar and circumstellar
  environments. However, several aspects regarding C$_{60}$ in space
  are not well understood yet, such as the formation and excitation
  processes, and the connection between C$_{60}$ and other
  carbonaceous compounds in the interstellar medium, in particular
  polycyclic aromatic hydrocarbons (PAHs). In this paper we study
  several photodissociation regions (PDRs) where C$_{60}$ and PAHs are
  detected and the local physical conditions are reasonably well
  constrained, to provide observational insights into these
  questions. C$_{60}$ is found to emit in PDRs where the dust is cool
  ($T_d = 20-40$~K) and even in PDRs with cool stars. These results
  exclude the possibility for C$_{60}$ to be locked in grains at
  thermal equilibrium in these environments. We observe that PAH and
  C$_{60}$ emission are spatially uncorrelated and that C$_{60}$ is
  present in PDRs where the physical conditions (in terms of radiation
  field and hydrogen density) allow for full dehydrogenation of PAHs,
  with the exception of Ced~201. We also find trends indicative of an
  increase in C$_{60}$ abundance within individual PDRs, but these
  trends are not universal. These results support models where the
  dehydrogenation of carbonaceous species is the first step towards
  C$_{60}$ formation. However, this is not the only parameter involved
  and C$_{60}$ formation is likely affected by shocks and PDR age.
\end{abstract}

\section{INTRODUCTION}

Buckminsterfullerene or C$_{60}$ was first discovered in laboratory
experiments aimed to understand the spectroscopy of carbon chain
molecules in the interstellar medium (ISM) and circumstellar envelopes
\citep{kro85}. Based on their study, the discoverers of C$_{60}$
concluded that the molecule corresponded to a new form of carbon
organized as a truncated icosahedron, usually compared with the old
style black and white ``soccer'' ball.

C$_{60}$ is a super-stable molecule and is considered the prototypical
fullerene, cage-like molecules of pure carbon. \citet{kra90} developed
a method to produce bulk quantities of C$_{60}$ by evaporating
graphite electrodes in a helium atmosphere. They also confirmed the
structure of the molecule through X-ray diffraction and its infrared
(IR) spectrum. The discovery of fullerenes has also opened the
research of carbon nanotubes, another different form of carbon which
combines the properties of graphite and fullerenes \citep{iji91}.

The values for the IR active modes are 526, 575, 1182 and
1429~cm$^{-1}$, which correspond to 18.9, 17.4, 8.5 and 7.0~$\mu$m
\citep{men00}. While the values for the frequencies show a good
agreement, reported intrinsic band strengths vary widely. Several
theoretical and experimental works have been carried out
\citep{cha92,fab96,cho00,igl11}, but the results have large
differences.

From its very discovery, C$_{60}$ was considered a potential component
for interstellar dust and speculated to be the carrier of some of the
diffuse interstellar bands (DIBs). Its survival in ISM conditions is
supported by its high stability. However, until recently, unequivocal
detection has not been possible. A first tentative detection was the
association of two, weak, far-red absorption bands with C$_{60}^+$
\citep{foi94}. This identification has been contested and the issues
were never fully resolved \citep{mai94}. The main obstacle for the
detection of fullerenes in emission stems from the fact that the
mid-IR spectra of almost all interstellar sources are dominated by the
vibrational spectrum of polycyclic aromatic hydrocarbons
(PAHs). Because of this, in sources with strong PAH emission, small
amounts of fullerenes are difficult to detect, with their emission
hidden by the PAH bands and the continuum.

\citet{cam10} recognized C$_{60}$ and C$_{70}$ bands in the spectrum
of the planetary nebula (PN) Tc~1, which has no strong PAH
bands. After the detection of these transitions and their association
with C$_{60}$ several more objects have been investigated, showing
that C$_{60}$ exists in a variety of sources with different
evolutionary stages and physical conditions. Most of these works have
dealt with PNe, including galactic and extragalactic sources in the
Magellanic Clouds \citep{gar11,ots12}, proto-PN \citep{zha11} and in
circumstellar envelopes of binary post-asymptotic-giant-branch stars
\citep{gie11}. There has also been a detection towards a young stellar
object (YSO) and a Herbig Ae/Be \citep{rob12}, which represent
isolated pre-stellar objects. C$_{60}$ was also detected in
photodissociation regions (PDRs), associated with both, reflection
nebula \citep[RN;][]{sel10} and H {\sc ii} regions \citep{rub11}.

The excitation mechanism of C$_{60}$ is not yet clear. This is an
important question to address since it determines which physical
conditions of its environment are traced by the bands. Two different
mechanisms have been suggested: \citet{cam10} consider the band ratios
from Tc~1 to be consistent with a thermal distribution of the excited
states, deriving an excitation temperature of $\sim 330$~K. This
mechanism requires that C$_{60}$ is not in the gas phase, but in solid
state or deposited on dust grains.

Another excitation mechanism, proposed by \citet{sel10}, assumes that
C$_{60}$ remains in the gas phase and the bands originate from IR
fluorescence. This is widely accepted as the excitation mechanism for
PAHs and consists in the absorption of a single UV photon, which
leaves the molecule highly excited and leads to a redistribution of
the energy between the vibrational modes. While for NGC~7023 the
reported 7.0/18.9um ratio is in good agreement with fluorescence
models, a much lower value was reported for this ratio in the
reflection nebula NGC~2023 \citep{sel10}. We note though that this
ratio is very difficult to determine in spectra dominated by the PAH
features as the underlying plateau emission is very strong and
broad. This holds for both NGC 7023 and NGC 2023 and is compounded
when the program {\sc pahfit} is used to extract the intensity for
highly blended, weak features as the adopted intrinsic Lorentzian
profile \citep[inappropriate for fluorescence from highly excited
species, cf.][]{tie08} locates much of the emission in ill-fitted
wings. Moreover, thermal as well as fluorescence analysis of observed
C$_{60}$ band ratios in space are hampered by the unknown intrinsic
strength of these modes.

Recently, \citet{ber13} reported a detection of C$_{60}^+$ in NGC~7023
through four bands at 6.4, 7.1, 8.2 and 10.5~$\mu$m. This would
support the idea that C$_{60}$ is in the gas-phase, at least in
environmental conditions similar to those found in NGC~7023.

The formation of C$_{60}$ is also the subject of debate. It seems
self-evident to consider that C$_{60}$ could be build-up from small
(hydro)carbon chains in the C-rich ejecta of Tc~1 and other PNe whose
spectra are dominated by C$_{60}$. In H-poor environments (not the
case of these PNe), simple formation paths analogous to the chemical
routes described by \citet{kro85} might lead to C$_{60}$ formation
\citep{che00}. However, neither mechanism is efficient enough to
account for the widespread detection of C$_{60}$ in the ISM.

Another hypothetical formation route in PNe is through photo-chemical
processing of hydrogenated amorphous carbon
\citep[HAC;][]{bern12,mic12}. The details of this mechanism require
the presence of large HAC clusters ($N_\mathrm{C} > 100$), with a high
H atomic fraction (i.e.\ not exposed to strong or continued UV
radiation). The sudden exposure of these small compounds to UV photons
is speculated to lead to dehydrogenation and aromatization. This would
result in giant fullerene cages which will shrink to C$_{60}$ and
C$_{70}$ through C$_2$ losses.

Finally, \citet{ber12} recognized that the abundance of C$_{60}$
increases rapidly near the illuminating star of the reflection nebula
NGC~7023, while the abundance of PAHs decreases. They propose that UV
processing of PAHs leads to fullerene formation. Under the influence
of radiation PAHs will first become dehydrogenated, leading to the
formation of graphene sheets. A continued exposure to high energy
photons is expected to remove carbon from within the structure, which
will force the formation of pentagonal rings and the bending of the
sheet. This process of destruction might lead to several different
intermediate stages. Because of its high stability, C$_{60}$ is
expected to be the photo-product with the longest lifetime, and
therefore it can be present at $\sim 0.01$\% of the elemental C near
the star.

In this work, our goal is to contribute to the understanding of the
origin and evolution of C$_{60}$ in environments where the presence of
PAHs is also strong. We test the formation mechanism proposed by
\citet{ber12} by comparing the variations of the abundance of C$_{60}$
and PAHs as a function of physical conditions ($G_0$,
$n_\mathrm{H}$). We also consider the spatial variations of these
abundances within the sources, in particular with respect to the
position of strong sources of radiation.  We used data from IRS
\citep{hou04} on-board of the {\it Spitzer Space Telescope}
\citep{wer04} and the PACS instrument \citep{pog10} on the {\it
  Herschel Space Observatory} \citep{pil10}, for three PDRs: NGC~2023
North, Ced~201 and RCW~49. We complement this analysis by including
observations of the Horsehead nebula, IC~63 and Orion's Veil. For
these latter sources we do not give a detailed analysis since there is
no detection of C$_{60}$, with the exception of Orion's Veil, which
was studied by \citet{boe12}.

This paper is organized as follows: in Sec.~\ref{sec:sources} we give
a description of the physical and chemical conditions of the sources,
taken from the literature. In Sec.~\ref{sec:data_red} we describe the
data from IRS and PACS that we use and explain the data reduction
process. In Sec.~\ref{sec:analysis} we present the reduced spectra and
feature maps obtained from IRS, as well as the derivation of the UV
field intensity from the PACS for each source. We also describe the
observed variations of C$_{60}$ and PAH abundances within each
source. In Sec.~\ref{sec:disc} we discuss our results and compare them
with the C$_{60}$ formation model of \citep{ber12}. Finally, in
Sec.~\ref{sec:concl} we summarize the main results and conclusions of
this work.

\section{SOURCES}\label{sec:sources}
\subsection{NGC~2023}\label{sec:ngc2023}

NGC~2023 is at a distance of $330-385$~pc and is illuminated by
HD~37903. This star is usually classified as a B1.5~V star
\citep{bro94}, although more recent works have concluded that it is a
Herbig Ae/Be star of spectral class B2~Ve \citep{moo09}, based on UV
spectra and IR excess. This RN is part of the dark cloud L~1630 in
Orion, which also hosts the famous Horsehead nebula (B33). Surrounding
HD~37903 there is a small H {\sc ii} region, with a size of 0.015~pc
\citep{kna75,pan76}. A far-IR study by \citet{har80}, showed that the
molecular cloud L~1630 lies behind HD~37903. The age of the stars
  in NGC~2023 lies in the range $1-7$~Myr, with the most massive stars
  falling towards the lower range \citep{lop13}.

\begin{figure}[t!]
  \centering
  \includegraphics[width=\columnwidth]{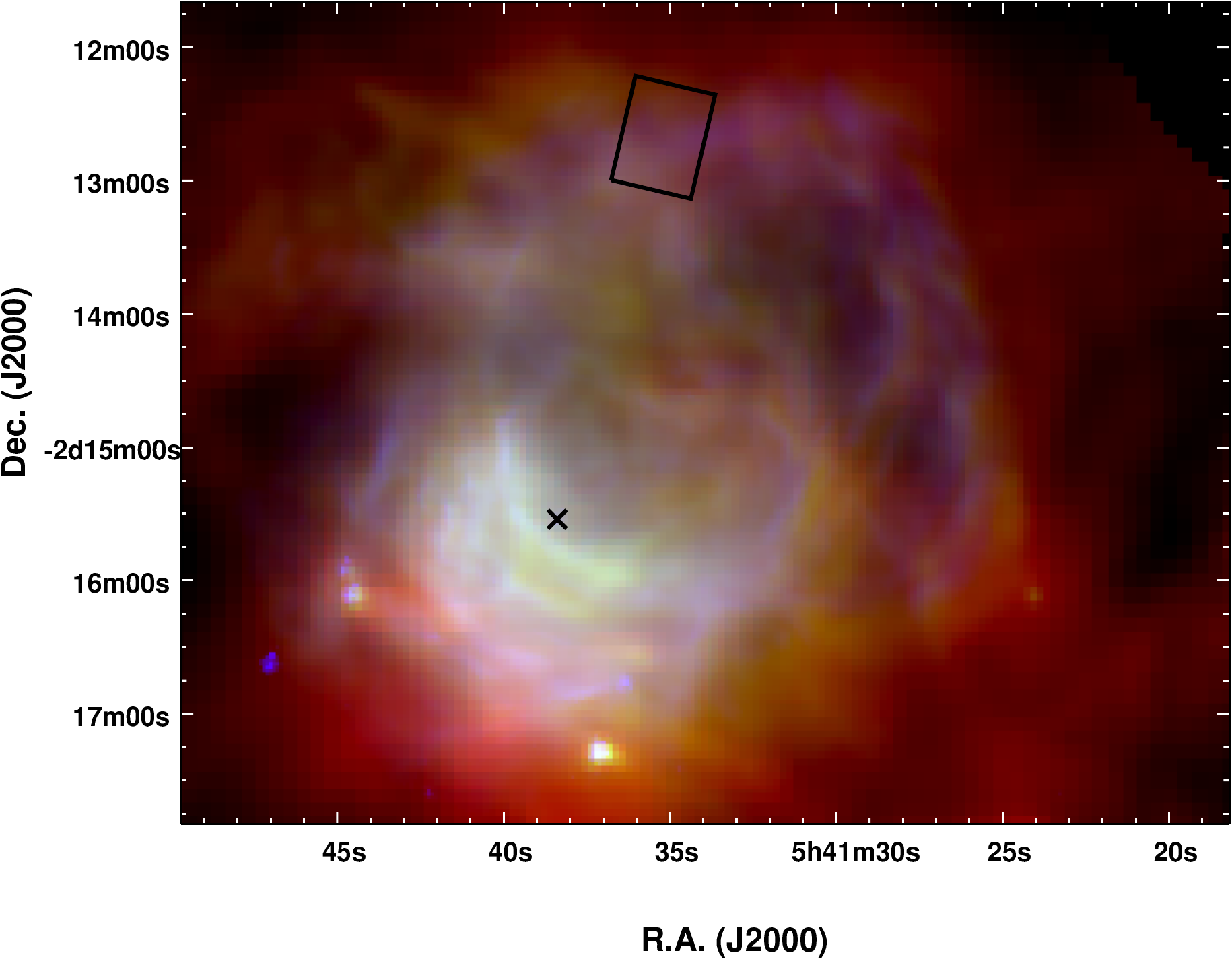}
  \caption{Composite image of NGC~2023. In blue is the IRAC 8~$\mu$m
    image, showing the distribution of PAHs. In green is the PACS
    70~$\mu$m map, which traces warm dust grains. In red is PACS
    160~$\mu$m, tracing colder dust. The cross indicates the position
    of HD~37903 and the black box shows the IRS field that we will use
    for this work.}
  \label{fig:ngc2023}
\end{figure}

Fig.~\ref{fig:ngc2023} gives a general overview of NGC~2023. We can
see that the shape of the reflection nebula is roughly circular, but
HD~37903 is not completely centered in it but displaced to the
southeast. The radiation from HD~37903 is the main source of UV
photons and creates a bubble-shaped cavity, particularly clear in the
IRAC 8~$\mu$m image (Fig.~\ref{fig:ngc2023}), which traces mostly PAH
emission. Outside the cavity most of the PAHs become faint as expected
because of the extinction of the UV field. Surrounding the cavity,
several filaments or ridges can be observed, most clearly in PAH and
warm dust emission. In particular, close to HD~37903, the southern
ridge is the most luminous part of the nebula, hosting a high
concentration of YSOs \citep{lad91,moo09}. Other less luminous ridges
can be seen to the north and the west of HD~37903. The black box in
Fig.~\ref{fig:ngc2023} shows the IRS field that will be the focus of
our analysis, which lies on top of the northern ridge of NGC~2023.

From observations of far-IR fine structure lines of [C~{\sc ii}],
[O~{\sc i}] and [Si~{\sc ii}], \citet{ste97} have established that the
environment in NGC~2023 is clumpy. Their observations of the northern
PDR are not spatially coincident with the position of the IRS field we
analyze, corresponding to a position closer to the HD~37903. For this
position they fit a model with $G_0 = 2\times 10^4$ and $n_0 = 3\times
10^3$~cm$^{-3}$. \citet{pil12} give a model for the same IRS field we
observe. They derive $n_0$ as a function of the distance to HD~37903
fitting a power law for the density profile, from the density in the
diffuse region (a free parameter in their fit) up to a maximum
density, which they take from \citet{fue95}, based on CN
observations. The density ranges from $2.4~\times 10^3$~cm$^{-3}$ to
$2\times 10^4$~cm$^{-3}$ throughout the field.

\begin{figure}[t!]
  \centering
  \includegraphics[width=\columnwidth]{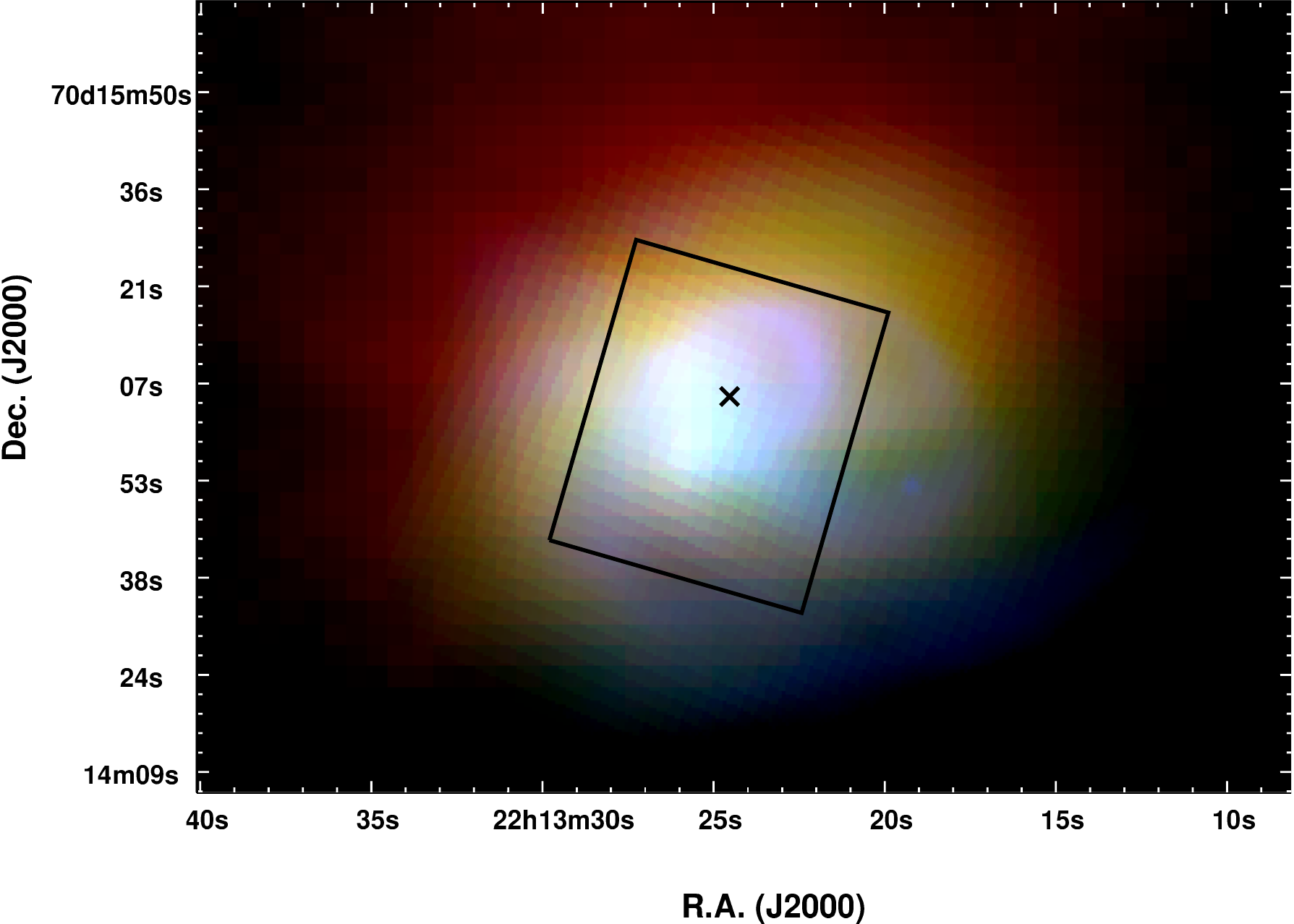}
  \caption{Composite image of Ced~201. Blue corresponds to the IRAC
    8~$\mu$m image, green to MIPS 24~$\mu$m and red to PACS
    70~$\mu$m. The box is the IRS field that will be used in this
    work. The cross marks the position of the illuminating star
    BD~+69~1231.}
  \label{fig:ced201}
\end{figure}

\citet{sel10} reported the detection of three features of C$_{60}$ at
7.0, 17.4 and 18.9~$\mu$m in NGC~2023. A recent work by \citet{pee12}
also detected extended C$_{60}$ emission through the 18.9~$\mu$m
feature in two locations inside the nebula, using the IRS instrument
on board of {\it Spitzer}. They found in their southern position that
the observed spatial variations of C$_{60}$ and PAHs are consistent
with \citet{ber12}, with C$_{60}$ appearing closer to the position of
HD~37903 than the PAH features. However, in their northern position
they find the opposite, with C$_{60}$ peaking further away from the
illuminating star than the PAHs, which they suggest may be due to
geometrical effects.

\subsection{Ced~201}\label{sec:ced201}

Ced~201 is an unusual RN, since it is the result of a chance encounter
of a molecular cloud with a runaway B9.5~V star (BD~+69~1231). This
was estimated considering a radial velocity difference between the
cloud and the molecular cloud of $\sim 12$~km~s$^{-1}$
\citep{wit87}. In most RNe the illuminating star is formed within the
cloud. Due to its velocity, the star in Ced~201 not only affects the
molecular cloud through the radiation field, but also induces shocks
and turbulence. An arc-like structure can be observed to the east of
the star, particularly clear in white in Fig.~\ref{fig:ced201}. This
structure has been interpreted as a shock due to the velocity
difference of the star and the molecular cloud. However, no signature
of a shock has been observed in the line profiles of CO ($2-1$)
\citep{ces00}. The molecular cloud is part of the Bok Globule B175
(L~1219), at a distance of $\sim 400$~pc \citep{cas91}.

Fig.~\ref{fig:ced201} shows the region of Ced~201. The star is
entering the molecular cloud from the west, generating the
aforementioned arc-like structure . We can see that the large dust
grains (in red) and with them the molecular cloud itself, are located
to the top of the image. The hot dust, traced by MIPS 24~$\mu$m, is
located closer to BD~+69~1231. PAH emission, traced by IRAC 8~$\mu$m
in Fig.~\ref{fig:ced201} (blue), is seen mostly located near the
  star and to the south. PAHs and large dust grains coexist in the
proximity of BD~+69~1231.

Given that the PDR is the product of a chance encounter, we can
estimate its age by the measured velocity of the star. In order to do
this we use the measured proper motion of the star \citep{hog00} and
we calculate the angular distance from the star to the edge of the
cloud in the direction of the proper motion. This gives us a rough
estimate of $\sim 1500$~yr.

\citet{you02} used far-IR fine fine structure line intensities of
[C~{\sc ii}] and [O~{\sc i}] (for the latter only upper limits were
derived) to determine that the gas physical conditions are $n =
4\times 10^2$~cm$^{-3}$ and $T = 200$~K, while for the radiation field
they find $G_0 \sim 300$. On the other hand a previous study by
\citet{kem99} found higher values $n = 1.2\times 10^{4}$~cm$^{-3}$ and
$T \sim 330$~K. They also find $n(\mathrm{H}_2) = (5\pm 1)\times
10^3$~cm$^{-3}$. These last values are derived from a PDR model that
fits the observed fine structure atomic lines (again [C~{\sc ii}] and
upper limits of [O~{\sc i}]) and molecular sub-mm lines. They detected
CO, $^{13}$CO and HCO$^+$ and in addition put upper limits for CS and
C$^{18}$O. The discrepancy in the values for the density stems from
the fact that that \citet{kem99} based their model not only on the
fine-structure lines but also on molecular data. We expect that their
values represent a better constraint on the density values. This is
also supported by \citet{cas91}, who finds $n > 2000$~cm$^{-3}$ based
on dust emission from far-IR.

\subsection{RCW~49}\label{sec:rcw49}

RCW~49 is one of the most luminous and massive H {\sc ii} regions in
the galaxy. The region is powered by the compact cluster Westerlund~2
\citep{wes60} where several OB stars have been detected, including at
least twelve O-stars earlier than O7, the earliest of type O3~V((f))
\citep{rau07}. It also hosts two Wolf-Rayet stars, WR20a and
WR20b. WR20a was found to be a binary system of two WN6ha stars with
individual masses of $\sim 70$~$M_\odot$ \citep{rau04}. WR20b appears
to be a single star with spectral type WN6ha.  The presence of
Wolf-Rayet stars imply a cluster age of $2-3$~Myr \citep{pia98}. The
molecular density and kinetic temperature have been determined through
$^{12}$CO, and $^{13}$CO observations by \citet{oha10}, with the
temperature ranging from 30 to 150~K and a typical density of
$n_{\mathrm{H}_2} \sim 3000$~cm$^{-3}$. Y.\ Sheffer et al.\ (in
preparation) estimate the density in five of the IRS fields that we
consider in this work. They fit models to the H$_2$ IR line
intensities in order to get the density parameters. These values will
be discussed in detail in Sec.~\ref{sec:nH_estimate}.

\begin{figure}[t!]
  \centering
  \includegraphics[width=\columnwidth]{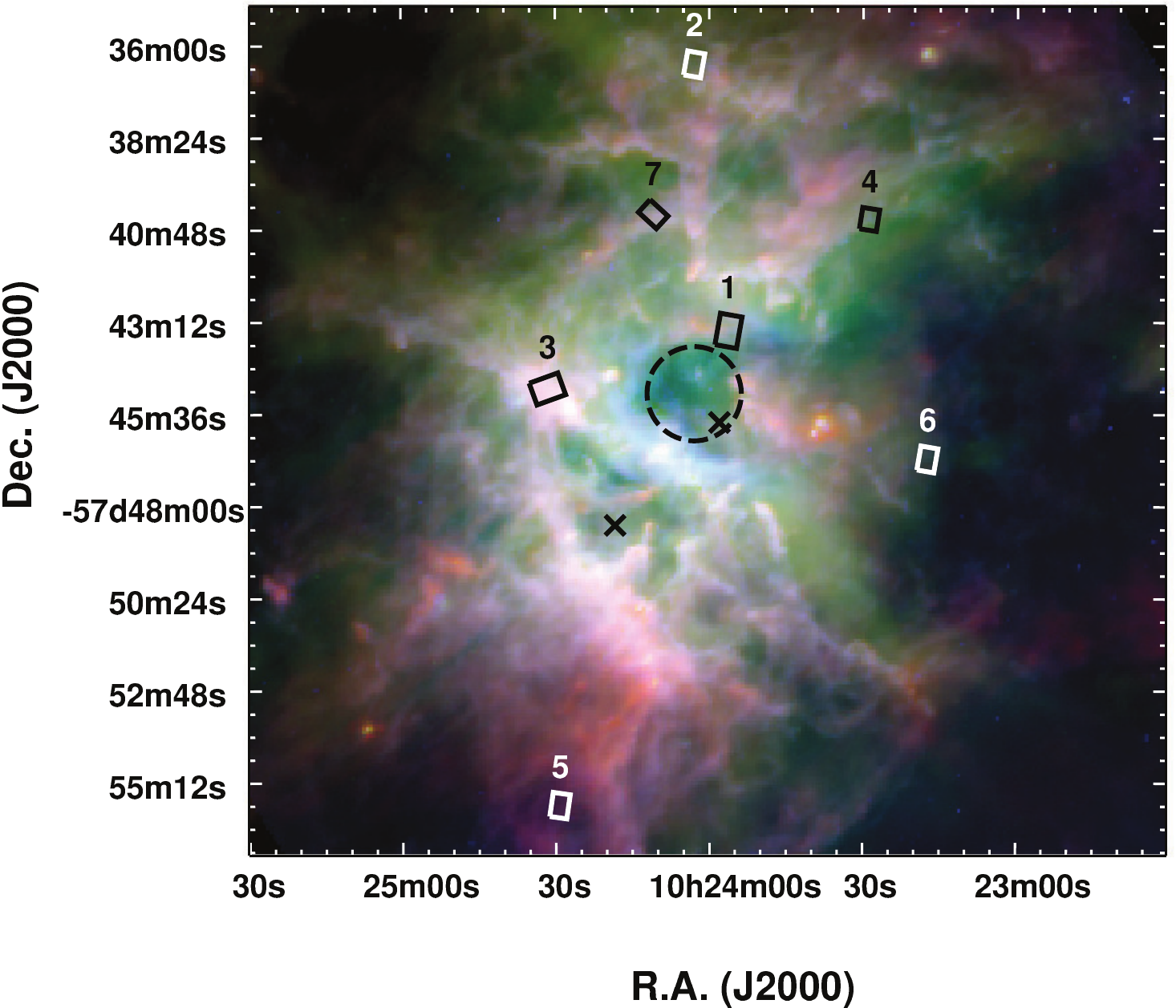}
  \caption{Composite image of RCW~49. Blue corresponds to the IRAC
    8~$\mu$m image, while green and red correspond the PACS images at
    70~$\mu$m and 160~$\mu$m respectively. The boxes show the seven IRS
    fields which will be used in this work. The numbers on top of each
    box correspond to the nomenclature used from here on to refer to
    the individual regions. The crosses indicate the position of both
    Wolf-Rayet stars, with WR20a to the north and WR20b to the
    south. Finally, the dashed circle indicates the position of the
    cluster Westerlund~2, which hosts most of the OB stars in the
    region.}
  \label{fig:rcw49}
\end{figure}

Fig.~\ref{fig:rcw49} shows the environment of RCW~49. PAHs and dust
are seen all over the nebula, with the brightest ridges near the
Wolf-Rayet stars. The overall shape is somewhat elongated in the
north-south direction, resembling a kidney. Two prominent bubbles are
observed in radio continuum images, one surrounding Westerlund~2 and
the other around WR20b. The bubble around Westerlund~2 is open to the
west \citep{whi97}. In this region dust and PAHs coexist with the
ionized gas or are at least embedded in neutral gas mixed with the
ionized region \citep{chu04}. PAH emission, traced by IRAC 8~$\mu$m
(blue in Fig.~\ref{fig:rcw49}), is more intense in the surroundings of
Westerlund~2. The ridges, bubbles and pillars that can be seen in
Fig.~\ref{fig:rcw49} are evidence of the strong interaction of the
parental molecular cloud with the stellar radiation and winds from
Westerlund~2 and the more recently formed stars. As we see, the seven
IRS fields considered in this work cover regions with very different
conditions, allowing us to probe a large range of UV field
intensities. In the following, when we refer to individual positions
we will use the numbers shown above the respective box in
Fig.~\ref{fig:rcw49}.

The distance to RCW~49 has been a subject of debate, with different
values ranging from 2.3 to 7.9~kpc \citep{bra93,whi97,mof91}, where
the lower limit comes from radio continuum observations and kinematic
studies and the upper limit is derived from the luminosity distance. A
recent photometric analysis by \citet{car13} derives the reddening and
extinction law towards several members of Westerlund~2. With this,
they correct the apparent distance modulus in order to obtain the
distance to the cluster, $d = 2.85\pm 0.43$~kpc.

\subsection{Additional Sources}

We consider additional observations of Orion's Veil \citep{boe12} and
NGC~7023 \citep{ber12}, for which C$_{60}$ has been detected. In the
case of Orion's Veil there are 11 positions available, which are
labeled with increasing distance to Orion's Bar as I4--1, M1--4 and
V1--3 respectively. From these we do not use the two closest regions,
I4 and I3, given that in the corresponding PACS images they suffer
from strong contamination with the emission from the Bar. We also
discard the farthest region, V3, considering that it looks edge on to
the back PDR, and thus does not probe exactly the same environment as
the remaining regions \citep{boe12}.

We also considered for our results and discussion two PDRs for which
the C$_{60}$ is not detected: IC~63 and the Horsehead nebula. In the
case of the Horsehead, we used data corresponding to the ``mane'',
where there is no detection of any excess at about 19~$\mu$m (where
the strongest C$_{60}$ feature falls) over the continuum, which is
strong in this region. We consider as upper limits the highest value
that could be fitted by {\sc pahfit} without creating a notorious bump
(always fitted as a detection below 3$\sigma$). For IC~63, there are
areas where emission over the continuum that cannot be fitted by the
[S{\sc iii}] line alone. These pixels are fitted by {\sc pahfit} with
confidence over 3$\sigma$. Upon visual inspection, however, we find
that the fit is of poor quality and we consider that C$_{60}$ is not
detected there. Because of this, we consider the fit as an upper limit
only. Physical conditions for these additional sources will be
provided in Secs.~\ref{sec:G0} and \ref{sec:nH_estimate}

\section{DATA}\label{sec:data_red}
\subsection{IRS Data}\label{sec:irs}

We used data from the Infrared Spectrograph \citep[IRS;][]{hou04}
on-board of the {\it Spitzer Space Telescope} \citep{wer04}. The
spectrograph consists of four modules, according to the wavelengths
and resolution, short wavelengths with low resolution (SL), long
wavelengths with low resolution (LL), short wavelengths with high
resolution (SH) and long wavelengths with high resolution (LH). 

For the low resolution modules the wavelength coverage goes from
5.2~$\mu$m to 38.0~$\mu$m for the combination of SL and LL and have a
spectral resolution, $R = \lambda/\Delta\lambda$, between $\sim 60$
and $\sim 130$. In the case of the high resolution modules, the
combined coverage of SH and LH ranges from 9.9~$\mu$m up to 37.2~$\mu$m,
with a spectral resolution $R\sim 600$.

We gathered the spectral cubes from the Spitzer Heritage
Archive\footnote{http://sha.ipac.caltech.edu}. For NGC~2023 and
Ced~201 we used observations from the program with PID 3512 (PI
C. Joblin). In the case of NGC~2023, we considered observations
performed with SL and LL modules with AOR keys 12014848 and 12011264,
respectively. For Ced~201 we selected again SL and LL modules, in this
case with AOR keys 11047936 and 11047680 respectively.

For RCW~49 we have 7 fields, all of them in SH module, from the
program with PID 20012 (PI M. Wolfire). The AOR keys of each of them
are 13812992, 13813248, 13813504, 13813760, 13814016, 13814272 and
13814528. These regions are not contiguous and will be referred to
according to the numbers shown in Fig.~\ref{fig:rcw49}.

The data reduction and cube generation was performed using the {\sc
  cubism} software \citep{smi07a}. The recommended data reduction
begins by taking a background when available. None of the observations
has a dedicated background observation. The low resolution modes have
``outrigger'' slits, that fall outside the area of interest. In the
case of Ced~201 this slit falls on an empty patch of sky so we used
this to do the background subtraction. In the case of NGC~2023 the
slit falls in regions with important emission from the same source
which leads us not to consider a background for this source. In this
case and in RCW~49 the lack of background should not be a very
important issue since both sources are very bright.

\subsection{PACS Data}\label{sec:pacs}

In order to have a measure of the FIR intensities we used photometric
data from the Photodetector Array Camera \& Spectrometer
\citep[PACS;][]{pog10} on-board of the ESA {\it Herschel Space
  Observatory} \citep{pil10}. In all three sources we retrieved the
available data at 70 and 160~$\mu$m from the Herschel Science
Archive\footnote{http://herschel.esac.esa.int/Science\_Archive.shtml}. The
observations of NGC~2023, Ced~201 and RCW~49 have respective OBSIDs of
1342227049/50, 1342196809/10 and 1342255009/10.

\section{ANALYSIS}\label{sec:analysis}

\subsection{Spectral Fit}

In order to fit the different features from the IRS spectra we used
the tool {\sc pahfit} \citep{smi07b}. This tool fits a stellar
continuum and several modified black bodies at different
temperatures. These black bodies are used to fit the mid-IR continuum,
but their temperatures do not represent the temperature of the dust,
since grains emitting at these wavelengths do not reach an equilibrium
with the radiation field, but are rather transiently heated to high
temperatures \citep{dra03}. It also considers several unresolved
atomic lines as well as H$_2$ rotational lines. Finally it fits
different PAH features and the 18.9~$\mu$m feature.

\begin{figure}[t!]
  \centering
  \includegraphics[width=\columnwidth,height=!]{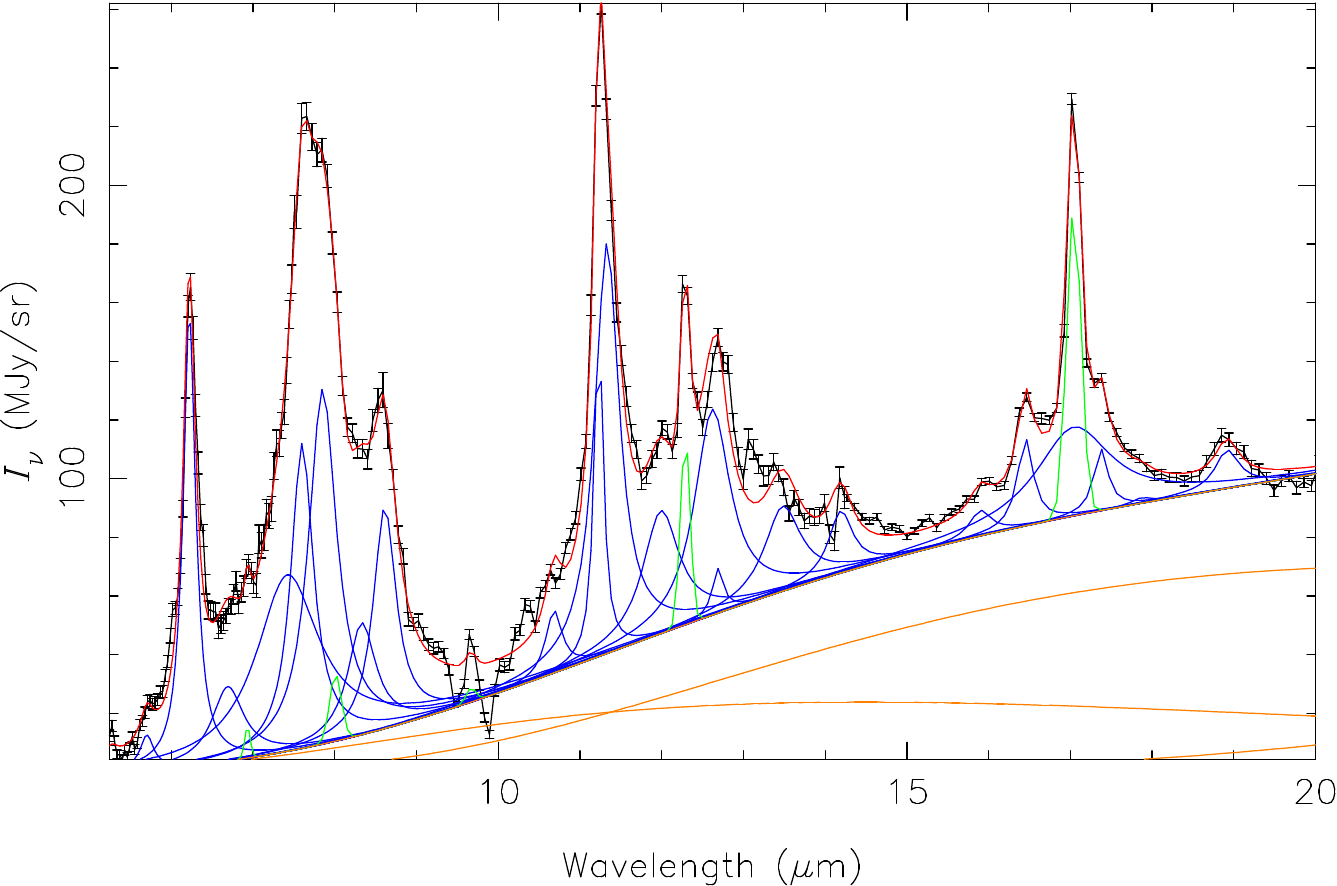}
  \caption{Examples of a fitted low resolution spectrum of
    NGC~2023. The black line with error bars is the observed spectrum
    while the red line is the total fitted spectrum. The individual
    components are shown in blue (PAH and C$_{60}$ features), green
    (unresolved H$_2$ lines) and orange (modified black body
    continua).}
  \label{fig:spectraL}
\end{figure}

\begin{figure}[t!]
  \centering
  \includegraphics[width=\columnwidth,height=!]{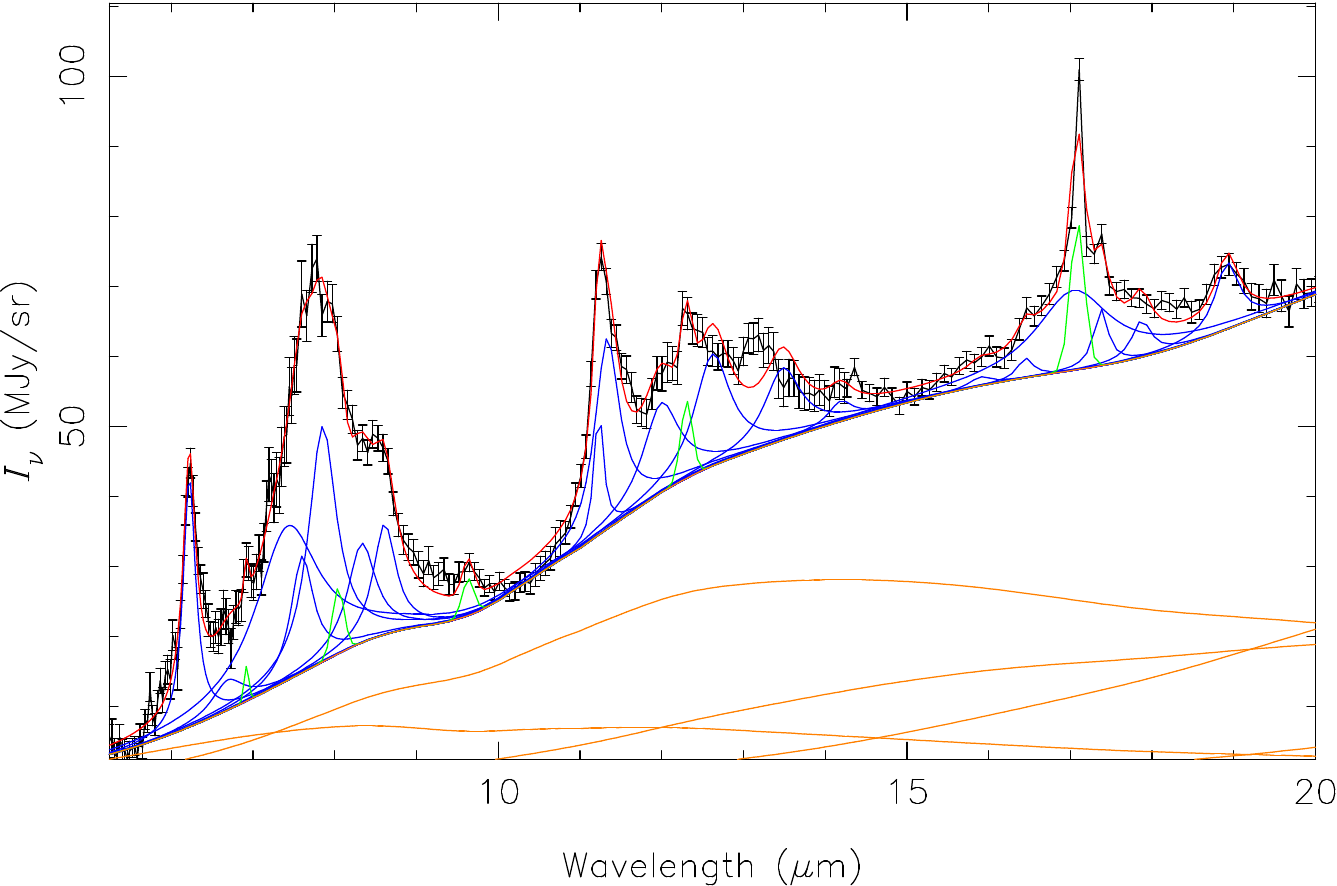}
  \caption{Example of a fitted low resolution spectrum from Ced~201.  The
    black line with error bars is the observed spectrum while the red
    line is the total fitted spectrum. The individual components are
    shown in blue (PAH and C$_{60}$ features), green (unresolved H$_2$
    lines) and orange (modified black body continua).}
  \label{fig:spec_ced201}
\end{figure}

In Fig.~\ref{fig:spectraL} we present an example of a low resolution
spectrum from NGC~2023 from the area where the 18.9~$\mu$m feature
peaks. The fit matches reasonably well the observed spectrum with the
exception of the PAH bands between 13 and 14~$\mu$m, for which the
peak positions seem to be somewhat different from the values
considered by {\sc pahfit}. However, these bands are rather weak and
when considering the integrated PAH band intensities their
contribution will be small. These characteristics are also observed in
the spectrum of Ced~201 (Fig.~\ref{fig:spec_ced201}). In both sources
the 18.9~$\mu$m feature associated with C$_{60}$ is clearly
present. The only unresolved lines fitted in these spectra are from
H$_2$ pure rotational lines.

The main PAH features at 6.2, 7.7, 8.6, 11.2, 12.7 and the plateau at
17.0~$\mu$m are also prominent. This allows us to consider in our
analysis a pixel by pixel comparison of the variations of C$_{60}$ and
PAHs. There are however some residual problems for the PAH bands: for
instance, in NGC~2023 there is a dip around 10~$\mu$m which may due to
saturation of the peak-up array as described in the IRS
handbook\footnote{http://irsa.ipac.caltech.edu/data/SPITZER/docs/irs}
(Sec.~7.3.5). In both sources the issues are not relevant for the
analysis since we will be focusing on the individual features as
fitted by {\sc pahfit} and no such feature falls in the wavelengths
with problems.

In the case of the 18.9~$\mu$m feature we found that, in some pixels,
the total intensity is underestimated. This defect is observed in
NGC~2023, for pixels that lie to the south of the field. The origin of
this problem comes from the continuum, which has a drop after
20~$\mu$m and thus is not properly fitted, affecting in turn the fit
of the 18.9~$\mu$m feature. In order to circumvent this issue and
check the accuracy of the fit in other pixels, we consider a local
linear continuum around 18.9~$\mu$m and later fit a Drude profile for
the feature. This procedure confirms that, the only significant
differences between the result of {\sc pahfit} and our fit for the
18.9~$\mu$m feature appear in the aforementioned region. For these
pixels, we use the result from our local fit instead of {\sc
  pahfit}. Other features in the area are not significantly affected
by the misplaced continuum.

One difference between the two sources is the feature-to-continuum
ratio. The PAH features in Ced~201 have a significantly smaller
intensity ratio with respect to the continuum than what we observe in
NGC~2023. This is not the case for the 18.9~$\mu$m C$_{60}$ feature,
which has a similar feature-to-continuum ratio in both sources. While
hard to disentangle from the H$_2$ $S(1)$ line and the 17.0~$\mu$m
plateau, the 17.4~$\mu$m band also seems to have similar
feature-to-continuum ratio in both sources. This could indicate a
large contribution of C$_{60}$ to this interstellar band
\citep{sel10,ber12,pee12}.

The high resolution spectra were also analyzed using {\sc
  pahfit}. However, we have to take into account that {\sc pahfit} was
created and optimized for fitting low resolution spectra, such as the
ones from Ced~201 and NGC~2023. In the case of RCW~49, the spectra
were taken in SH mode for all the regions considered here. This forces
us to change the input parameters in {\sc pahfit}. Moreover, for the
unresolved atomic and molecular lines we changed the width of the
Gaussians to match the improved resolution of the SH mode. We also
modified some of the PAH features to improve the fit.

\begin{figure}[t!]
  \centering
  \includegraphics[width=\columnwidth,height=!]{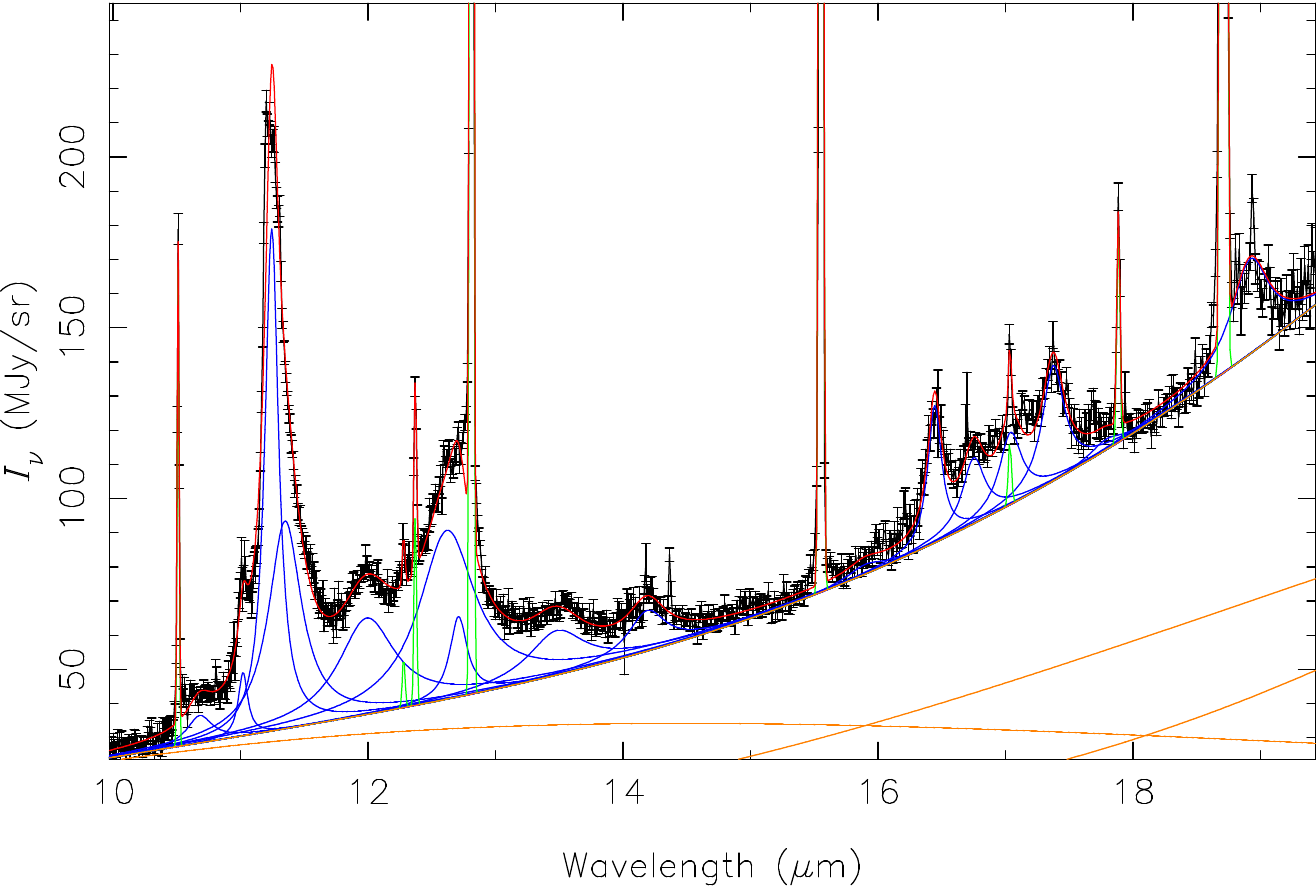}
  \caption{Example of fitted a high resolution spectrum from RCW~49,
    Region~2. The black line with error bars is the observed spectrum
    while the red line is the total fitted spectrum. The individual
    components are shown in blue (PAH and C$_{60}$ features), green
    (unresolved H$_2$ and atomic lines) and orange (modified black
    body continua).}
  \label{fig:spec_rcw49}
\end{figure}

\begin{figure*}[t!]
  \centering
  \includegraphics[width=.81\textwidth,height=!]{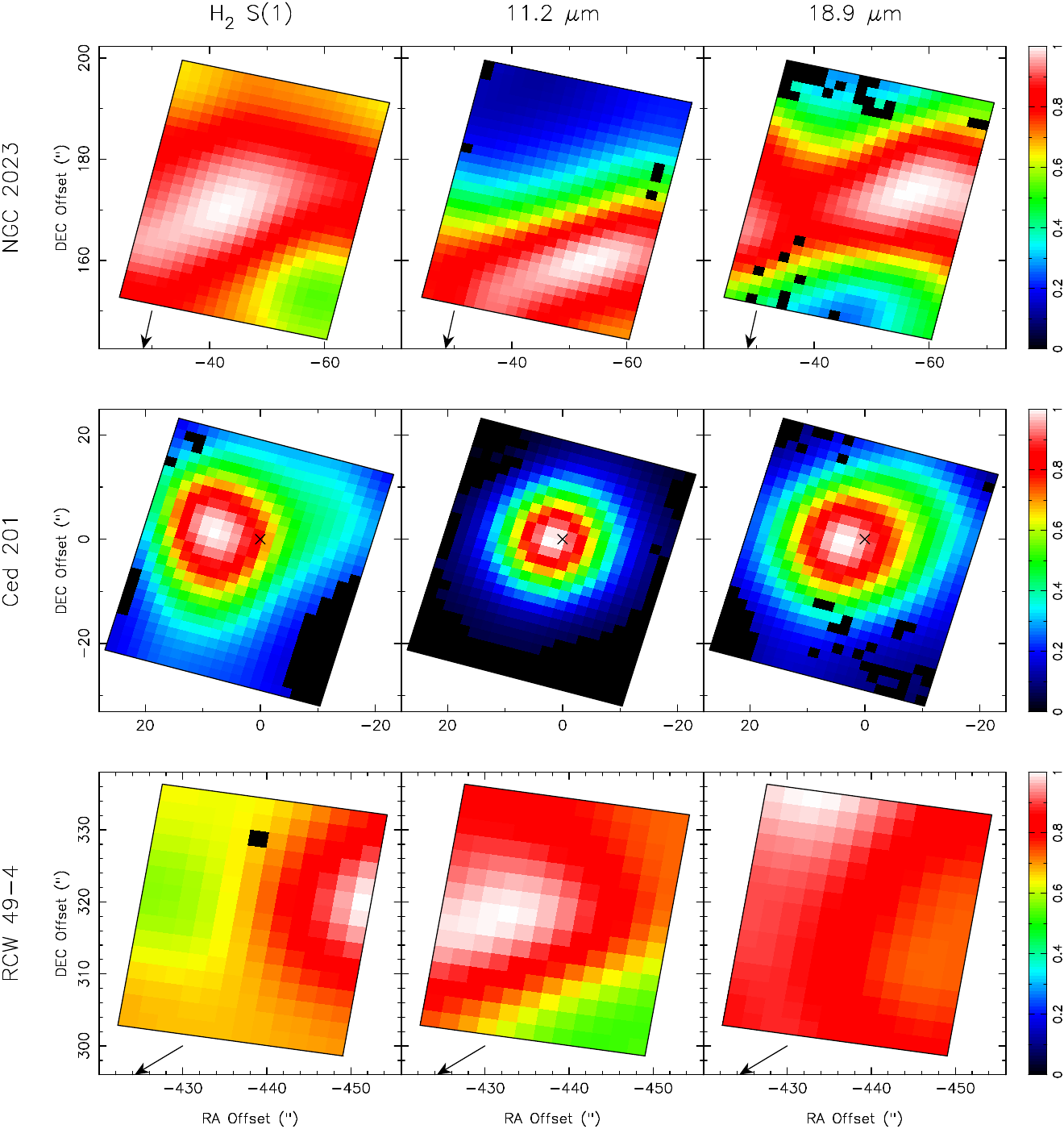}
  \caption{Intensity maps of the selected features: H$_2$ 0-0 $S(1)$
    line, 11.2~$\mu$m PAH feature and 18.9~$\mu$m C$_{60}$ feature,
    all convolved to the PACS 70~$\mu$m beam. The top panels
    correspond to NGC~2023, the middle ones to Ced~201 and the bottom
    panels to Reg.~4 in RCW~49. The color scale is normalized with
    respect to the maximum intensity in the corresponding frame. The
    offsets are with respect to the position of the illuminating star,
    towards which also the arrow or cross points (for RCW~49 we chose
    the position of WR20a).}
  \label{fig:sample_maps}
\end{figure*}

The fitting procedure follows the same principle as in the low
resolution mode. Besides the previously mentioned modifications of the
parameters we were also forced to remove all features that fall
outside the range covered by the SH mode since this causes problems
with the fitting routine. If this last modification is not implemented
then the wings of the adopted Drude profiles of the out-of-range PAH
features are sometimes fitted as part of the continuum, giving wrong
fits.

We also found that this modified fit runs into problems when the
feature to continuum ratio becomes small. This is the case for
region~1 in RCW~49. In particular at longer wavelengths the fit tends
to either over- or underestimate the continuum, resulting in poor fits
for the 18.9~$\mu$m feature. Furthermore, the 18.9~$\mu$m feature is
not visually recognizable in the spectra. However, the continuum at
this wavelength is very high for this region. Even the presence of an
18.9~$\mu$m feature with an intensity comparable with the highest
values detected among our sources would not be recognizable. For these
reasons we decided to exclude region~1 from our analysis.

We searched for potential errors in the fits for the continuum but in
all cases the fit is consistent with the data. In
Fig.~\ref{fig:spec_rcw49} we present an example of region~2 in RCW~49,
which has the strongest 18.9~$\mu$m feature. The 18.9~$\mu$m feature
varies in intensity between the different regions, but can be
recognized in the spectra in most pixels. In some cases the detection
is only marginal, but in our final analysis we exclude all points with
detection below 3$\sigma$ with respect to the error bars given by
{\sc pahfit}.

\subsection{Spectral Maps}

Once the spectrum for each pixel have been fitted we build maps for
each of the PAH features, the spectral lines and the 18.9~$\mu$m
feature. We also create maps for the errors, as derived by {\sc
  pahfit}. These maps give the integrated intensity for each spectral
feature and will be the basis for our analysis. In order to restrict
our analysis to those points with clear detection we also flagged
pixels for which the fitted intensity falls below a $3\sigma$ noise
level. Finally, in order to get smoother images and correct for cases
in which particular pixels have large differences in flux with respect
to its neighbors, we performed a median filter with a $2\times 2$
kernel.

In Fig.~\ref{fig:sample_maps} we present the maps of the 11.2~$\mu$m
PAH feature, the 18.9~$\mu$m feature corresponding to C$_{60}$ and the
H$_2$ 0-0 $S(1)$ line at 17.0~$\mu$m. We show the results for NGC~2023,
Ced~201 and Reg.~4 of RCW~49. We can see that in all the cases the PAH
feature peaks closer to the illuminating star than the C$_{60}$ or
H$_2$ lines. This contrasts with the observations on NGC~7023, where
C$_{60}$ peaks closer to the illuminating star than the PAHs.

In NGC~2023 we identify a bar-like structure in all the maps, running
from the southeast to the northwest. However, the exact position of
the bar varies: on the 11.2~$\mu$m map it is displaced to the south
when compared with the 18.9~$\mu$m or H$_2$ $S(1)$ maps. Other
variations observed in the bar concern the position of the peak
intensity, with 18.9~$\mu$m displaying two peaks located towards the
northwest and southeast of the bar, while H$_2$ $S(1)$ shows one
peak located roughly in the middle of the bar and 11.2~$\mu$m showing
also one peak, but displaced to the northwest.

In the case of Ced~201 there is a clear asymmetry in the cases of the
H$_2$ and C$_{60}$ maps. This asymmetry is also present but in a much
more contained way in the case of the PAH maps. In the case of H$_2$
and C$_{60}$, we can see in both maps that the arc towards the
northwest appears much more clearly than in the PAH maps. Also in
both cases, the peak intensity appears to the northeast of the star,
in the position of the main part of the molecular cloud as shown in
Fig.~\ref{fig:ced201}. In contrast, the PAH features are centered near
the star, and, for the 8.6 and 7.4~$\mu$m features, the coincidence
with IRAC 8~$\mu$m is very clear.

For RCW~49, we can see that the H$_2$ peaks normally avoid the peaks
of the PAH features, and are typically found further away from the
illuminating star. For the 18.9~$\mu$m feature, we can see that, in
most regions we have little variations in intensity. As is the case
for NGC~2023 and Ced~201, the 18.9~$\mu$m has a clearly different
spatial distribution when compared to the PAH features in all the
regions. In RCW~49, all the PAH features follow a similar pattern, with
minor variations in peak positions and relative strengths. However,
when comparing to the 18.9~$\mu$m maps, we find much stronger
variations in the peak position and relative intensities in most of
the sources. Sometimes the PAH and C$_{60}$ morphology show a
superficial resemblance, but the relative strength of the different
components is very different.

In summary, for all the sources where we detect the 18.9~$\mu$m
feature, its spatial distribution noticeably different with respect to
the distribution of the PAH features. Although the different PAH bands
show spatial variations with respect to each other, they resemble each
other much more closely than the 18.9~$\mu$m feature. In most cases,
the peak intensity of the C$_{60}$ band is seen farther from the
central star than the PAH bands. This behavior is the opposite of what
is observed in NGC~7023 \citep{ber12} and, in principle, would not
support the formation of C$_{60}$ by UV processed PAHs. It has been
suggested that the formation of C$_{60}$ in the ISM starts with the
dehydrogenation of PAHs which is a balance between UV photolysis and
reactions with H, we will consider in the next two subsections the
intensity of the local radiation field and the hydrogen density.

\subsection{Calculation of $G_0$}\label{sec:G0}

Considering that PAH and C$_{60}$ are stochastically heated by
UV-photons we will need the intensity of the UV field in order to have
a proper measure of the abundances. The intensity of the mid-IR bands
will be proportional to the product of the column density of PAHs or
C$_{60}$ with the UV field intensity.

We will determine the FUV field intensity from the FIR spectrum of
dust in terms of $G_0$, which is a measure of the FUV field in terms
of the Habing field \citep[$1.6\times
10^{-3}$~erg~cm$^{-2}$~s$^{-1}$;][]{hab68}.  It considers the flux of
photons with energies between 6~eV and 13.6~eV. One way of calculating
$G_0$ is to consider that in a PDR nearly all the radiation is
absorbed by dust in the UV and re-emitted in the far infrared
(FIR). Assuming this and that grains act like a modified black body
\citep[Sec.~9.7]{tie05} we can write,
\begin{equation}
  G_0 = 8.3\times 10^{3}\frac{V}{\ell S}\tau_{\nu_0}\nu_0^{-\beta}\int
  \nu^{\beta}B(\nu,T_d)\dif{\nu},\label{eq:g0}
\end{equation}
where $\nu_0$ is a reference frequency where the optical depth is
$\tau_{\nu_0}$, $\beta$ is the spectral index and $T_d$ is the dust
temperature. The factor $V/\ell S$ accounts for the geometry of the
cloud, $S$ is the surface area that is facing the illuminating star,
$V$ is the volume of the cloud and $\ell$ corresponds to the length
along the line of sight. The problem with this method is that the
three-dimensional geometry is generally not well known, introducing
uncertainties. For example, for an edge-on disk the geometric factor
would be equal to the ratio of the disk thickness over $\ell$, while
for a sphere this factor is equal to one. Since this method relies on
the emission from dust, it will underestimate the value of $G_0$ in
regions devoid of dust. In the three PDRs considered here the
environment is dusty and the IRS fields are away from hard sources of
radiation, so we expect our $G_0$ estimate to provide a reasonable
approximation.

To calculate the total FIR intensity we use the PACS images at 70 and
160~$\mu$m from each source. In evaluating eq.~(\ref{eq:g0}), we have
corrected for the emission missing in the 70 and 160~$\mu$m filters by
fitting modified black-bodies to those data. For this, we first
matched the 70~$\mu$m images to the 160~$\mu$m resolution. With this we
then fitted a modified black-body to these two points, on a pixel by
pixel basis. The values to be fitted are the dust temperature $T_d$
and the optical depth $\tau_{\nu_0}$. We fixed $\nu_0$ at 1000~GHz and
took $\beta = 1.7$. The value of $\beta$ in the ISM is usually found
to be somewhere between one and two, with some dependency on the dust
temperature \citep{dup03}. It has been observed that for $T_d \lesssim
20$~K then $\beta \sim 2$, while for $T_d \gtrsim 30$~K a value of
$\beta \sim 1.5$ is found \citep{she09}. We chose an intermediate
value since in all our sources we expect dust with temperatures in
both regimes. We are not fitting $\beta$ directly since we only have
two points for our fit which is not enough to realistically fit three
parameters. To make the fit we used a Levenberg-Marquardt fitting
routine. With this fit, we derive the value of $T_d$ and
$\tau_{\nu_0}$, which we can later use to determine $G_0$.

\begin{figure}[t!]
  \centering
  \includegraphics[width=\columnwidth,height=!]{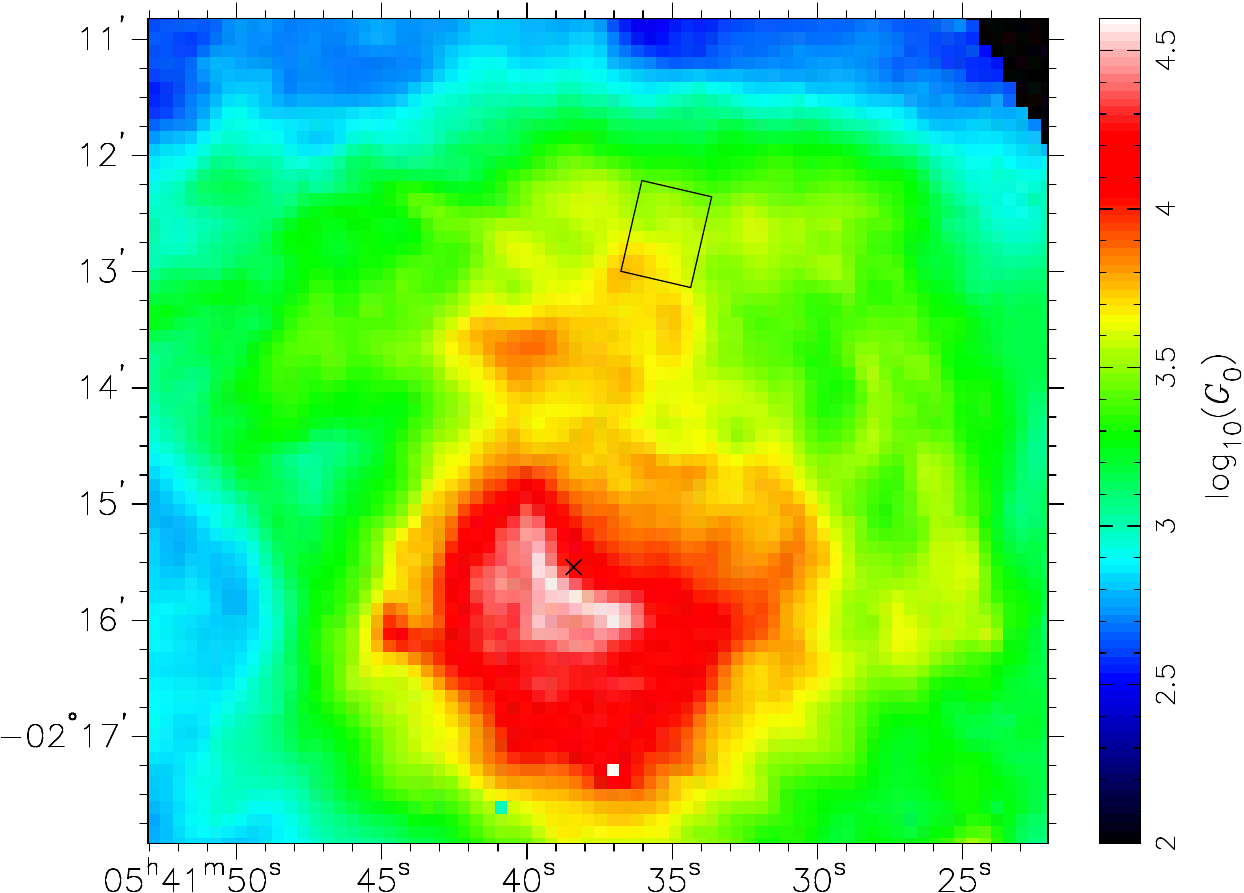}
  \caption{Map of $G_0$ calculated from the FIR intensity for
    NGC~2023 with an assumed a spherical geometry, thus setting
    $V/\ell S = 1$. The cross shows the position of the illuminating
    star and the square the position of the IRS field.}
  \label{fig:Go_ngc2023}
\end{figure}

We assumed spherical geometry in the three cases. For NGC~2023
(Fig.~\ref{fig:Go_ngc2023}) this is likely a good assumption since the
RN is almost circular, even though it is not exactly centered on the
illuminating star. An additional factor must be added in
eq.~(\ref{eq:g0}) to account for the fact that, particularly for B-type
stars, there is a significant part of the radiation at wavelengths
longer than 6~eV, that will heat the dust but do not correspond to the
FUV as defined by $G_0$. Based on the spectral type we use a factor of
0.7 for the fraction of the stellar luminosity falling in the range
between 6 and 13.6~eV \citep{ste97}. We check our calculation against
the result of \citet{she11}. They found $G_0 = 1.7\times 10^4$ in
NGC~2023~S, while the region we are focusing on is about 2.3 times
farther out. This is equivalent to a geometric dilution factor of 5.3,
giving an expected value of $G_0 = 3200$, which is in agreement with
the values seen in Fig.~\ref{fig:Go_ngc2023}.

\begin{figure}[t!]
  \centering
  \includegraphics[width=\columnwidth,height=!]{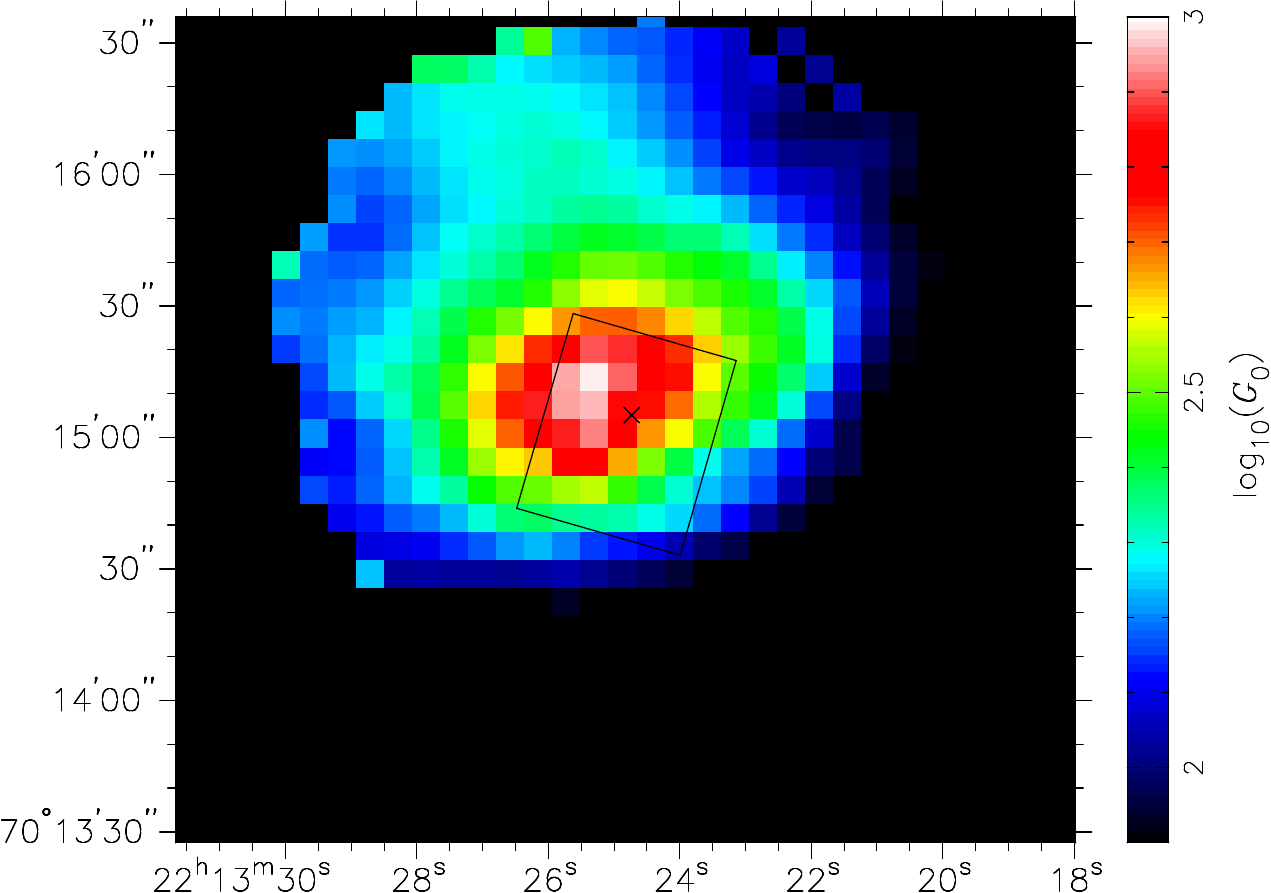}
  \caption{Map of $G_0$ calculated from the FIR intensity for Ced~201
    assuming spherical geometry ($V/\ell S = 1$). The cross shows the
    position of the illuminating star and the square corresponds to
    the IRS field.}
  \label{fig:Go_ced201}
\end{figure}

In the case of Ced~201 (Fig.~\ref{fig:Go_ced201}) the spherical cloud
assumption is not entirely realistic since we have a chance encounter
with the molecular cloud and the star is likely situated at the edge
of the cloud. We can see that in this case the peak of $G_0$ is
displaced from the illuminating star and, furthermore, the region with
highest $G_0$ corresponds to the arc structure seen northeast from
BD~+69~1231 in the molecular cloud, with much less intensity to the
southwest. In this case the fraction of the stellar luminosity
emitted between 6 and 13.6~eV corresponds to 0.3 \citep{you02}.

\begin{figure}[t!]
  \centering
  \includegraphics[width=\columnwidth,height=!]{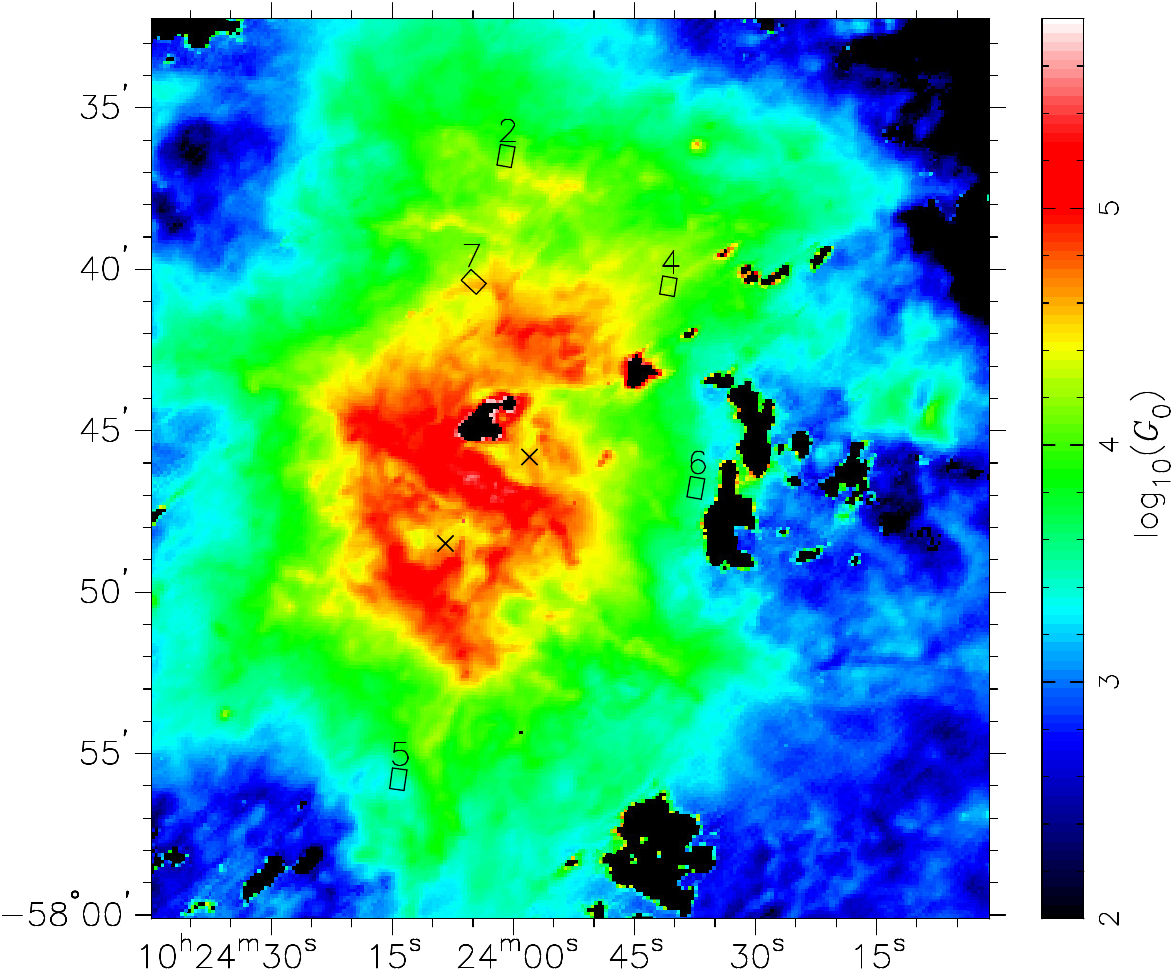}
  \caption{Map of $G_0$ calculated from the FIR intensity for RCW~49
    assuming spherical geometry ($V/\ell S = 1$). The crosses show the
    position of the illuminating stars and the squares the position of
    the five IRS fields for which we have density estimates.}
  \label{fig:Go_rcw49}
\end{figure}

In the case of RCW~49 (Fig.~\ref{fig:Go_rcw49}), the exact
three-dimensional geometry is not well known, but it looks
approximately circular and we expect that the spherical approximation
will not introduce large errors. We will take into account in our
discussion the possible effects that arise from changing the
geometrical factor. Since the illuminating source of RCW~49 is
dominated by the O-type stars we consider a factor of one for the
fraction of the luminosity emitted in the FUV. In reality, a
significant fraction will be emitted at wavelengths shorter than
13.6~eV in this case, but these photons will be absorbed and
downgraded in the H{\sc ii} region.

For the additional PDRs included in this work we only used the the
far-IR calculation of $G_0$ for Orion's Veil, resulting in $G_0$
ranging from $\sim 30000$ for the I2 field (the closest to the Bar) to
$G_0 \sim 3000-5000$ for the fields that lie farther from the Bar. For
NGC~7023 we used the values derived by \citet{ber12}, which result in
a lower estimate of $G_0 = 10000\pm 7000$ at the peak of the
18.9~$\mu$m feature. The Horsehead nebula was studied by
\citet{abe03}, who find $G_0 \sim 100$. For IC~63 the estimate by
\citet{jan94} is of $G_0 \sim 650$.

Finally, comparing Figs.~\ref{fig:ngc2023}--\ref{fig:rcw49} against
Figs.~\ref{fig:Go_ngc2023}--\ref{fig:Go_rcw49} we see that the general
structure seen in the composite images is clearly apparent in the
$G_0$ distribution. The $G_0$ maps show the most characteristic
features in the case of each PDR, such as the southern and northern
ridges in NGC~2023, the bright bar observed in RCW~49 between the
positions of WR20a and WR20b and the arc-like structure seen in
Ced~201 to the northeast of BD~+69~1231. The derived dust
temperatures inside the IRS fields of the three PDRs range between 20
and 40~K, which is much too low for emission from C$_{60}$ embedded in
grains.

\subsection{The Atomic H Density}\label{sec:nH_estimate}

In order to estimate the variations in abundance of PAHs and C$_{60}$
with respect to the physical conditions of each source we need to
derive the atomic hydrogen densities, $n_\mathrm{H}$, for each
region. We estimate these based on previous studies and we derive a
single value as representative for each of the regions we consider. A
summary of the results for each field is presented in
Table~\ref{tab:densities}.

For NGC~2023 we base our estimate on the densities calculated by
\citet{pil12}. Their model uses a varying density inside the IRS
field, which ranges from $2.4~\times 10^3$~cm$^{-3}$ to $2\times
10^4$~cm$^{-3}$. We will consider their value for the density in
the inner part of the field, which will dominate the emission.

In the case of Ced~201 we base our estimate of $n_\mathrm{H}$ on the
values for total hydrogen density and molecular densities from
\citet{kem99} and \citet{you02}. The $G_0$ value derived from the IR
measurement is in good agreement with the value derived from the
PDR/molecular cloud analysis of \citet{kem99}. As these authors also
included more reliable density tracers, we have adopted their value
for $n_\mathrm{H}$ and the IR derived value for $G_0$.

In RCW~49 we will use the values given by Y.\ Sheffer et al.\ (2014,
in preparation) for five out of the six IRS fields. They model the
H$_2$ rotational line emission in a manner similar to that in
\citet{she11}. The IRS SH observations cited in Sec.~\ref{sec:irs} are
used to derive the H$_2$ 0-0 $S(1)$ to $S(4)$. They additionally use
LL observations to cover the H$_2$ 0-0 $S(0)$ line. PDR models are
used to fit each field individually, giving H$_2$ column densities and
reference values of $n_0$. For a typical PDR, hydrogen in the surface
layer is predominantly atomic and we have $n_\mathrm{H} = n_0$. In
dense PDRs, however, the H/H$_2$ transition is pulled to the surface
and the atomic fraction varies from 1 to $10^{-3}$. For our sources,
this seems to apply in RCW~49 regions 2 and 4. For these two
positions, we hence adopted the atomic H fraction at $A_V = 0.5$ in
the detailed models of Y.\ Sheffer et al.\ (2014, in preparation);
$f_\mathrm{H} = 10^{-3}$ and $10^{-1}$ for regions 2 and 4
respectively. It should be realized that these values are very
uncertain.

The value derived for NGC~7023 by \citet{ber12} is of $n_\mathrm{H} =
150\pm 100$~cm$^{-3}$. For Orion's Veil, we estimate the density by
using the electron densities derived from [S {\sc ii}] optical lines
by \citet{rub11}. From their values we use a factor 20 to derive the
neutral density \citep{tie05}. In the Horsehead nebula, \citet{abe03}
estimate the density just behind the ionization front to be $8\times
10^3$~cm$^{-3}$. Finally, in the case of IC~63, we consider the
results from \citet{thi09}, who give a range from
$1000-5000$~cm$^{-3}$ for the PDR density.

\begin{deluxetable}{lccccr}[b!]
  \tablecaption{Average Physical Conditions and C$_{60}$
    Fractions.\label{tab:densities}} 
  \tablehead{ \colhead{Region} &
    \colhead{$n_\mathrm{H}$} & \colhead{$G_0$} &
    \colhead{$G_0/n_\mathrm{H}$} &
    \colhead{$f_{\mathrm{C},\mathrm{C}_{60}}$} & \colhead{Ref.} \\
    & \colhead{(cm$^{-3}$)} & & & ($10^{-5}$) & } 
  \startdata
  NGC~2023 & $2.4\times 10^3$ & $5200\pm 200$ & 2.1 & 4.2 & 1 \\
  Ced~201 & $1.2\times 10^4$ & $600\pm 60$ & 0.050 & 44 & 2 \\
  RCW~49-2 & $4.0\times 10^3$ & $11000\pm 1300$ & 2.7 & 13 & 3\\
  RCW~49-4 & $1.0\times 10^4$ & $20000\pm 3000$ & 2.0 & 7.2 & 3\\
  RCW~49-5 & $2.5\times 10^4$ & $3000\pm 400$ & 0.12 & 11 & 3\\
  RCW~49-6 & $2.5\times 10^4$ & $5100\pm 300$ & 0.21 & 12 & 3\\
  RCW~49-7 & $2.5\times 10^4$ & $34000\pm 2500$ & 1.4 & 5.5 & 3\\
  \hline
  NGC~7023 & $1.5\times 10^2$ & $10000\pm 7000$ & 60 & 9.9 & 4\\
  Veil-I2 & $8.7\times 10^3$ & $31000\pm 1900$ & 3.6 & 56 & 5,6\\
  Veil-I1 & $8.1\times 10^3$ & $22000\pm 500$ & 2.7 & 41 & 5,6\\
  Veil-M1 & $5.1\times 10^3$ & $15000\pm 600$ & 2.9 & 36 & 5,6\\
  Veil-M2 & $5.1\times 10^3$ & $13000\pm 600$ & 2.5 & 32 & 5,6\\
  Veil-M3 & $5.1\times 10^3$ & $8700\pm 400$ & 1.7 & 29 & 5,6\\
  Veil-M4 & $3.4\times 10^3$ & $5400\pm 200$ & 1.6 & 30 & 5,6\\
  Veil-V1 & $1.7\times 10^3$ & $3500\pm 100$ & 2.0 & 35 & 5,6\\
  Veil-V2 & $2.6\times 10^3$ & $5200\pm 100$ & 2.0 & 31 & 5,6\\
  Horsehead & $8.0\times 10^3$ & $\sim 100$ & 0.013 & $<4.3$ & 7\\
  IC~63 & $3.0\times 10^3$ & $\sim 650$ & 0.22 & $<6.0$ & 8,9
  \enddata
  \tablecomments{For NGC~2023, Ced~201, RCW~49 and the Veil, the
    values of $G_0$ were calculated following the procedure described
    in Sec.~\ref{sec:G0}. Other values of $G_0$ and $n_\mathrm{H}$ are
    taken from the given references. For NGC~7023 we use lower limits
    for $G_0$ and present the values at the peak of C$_{60}$
    abundance.}  \tablerefs{(1) \citet{pil12}; (2) \citet{kem99}; (3)
    Y.\ Sheffer et al.\ (in preparation); (4) \citet{ber12}; (5)
    \citet{boe12}; (6) \citet{rub11}; (7) \citet{abe03}; (8)
    \citet{jan94}; (9) \citet{thi09}.}
\end{deluxetable}

For all the PDRs considered here, the values of $n_\mathrm{H}$ carry
uncertainties that are difficult to estimate well. Since we are using
a single value as representative for each region, we have that the
variations of $G_0/n_\mathrm{H}$ within each field will be
representative of variations of $G_0$ only.

\subsection{Variations in Abundance}\label{sec:var_abund}

\begin{figure*}[t!]
  \centering
  \includegraphics[width=\textwidth,height=!]{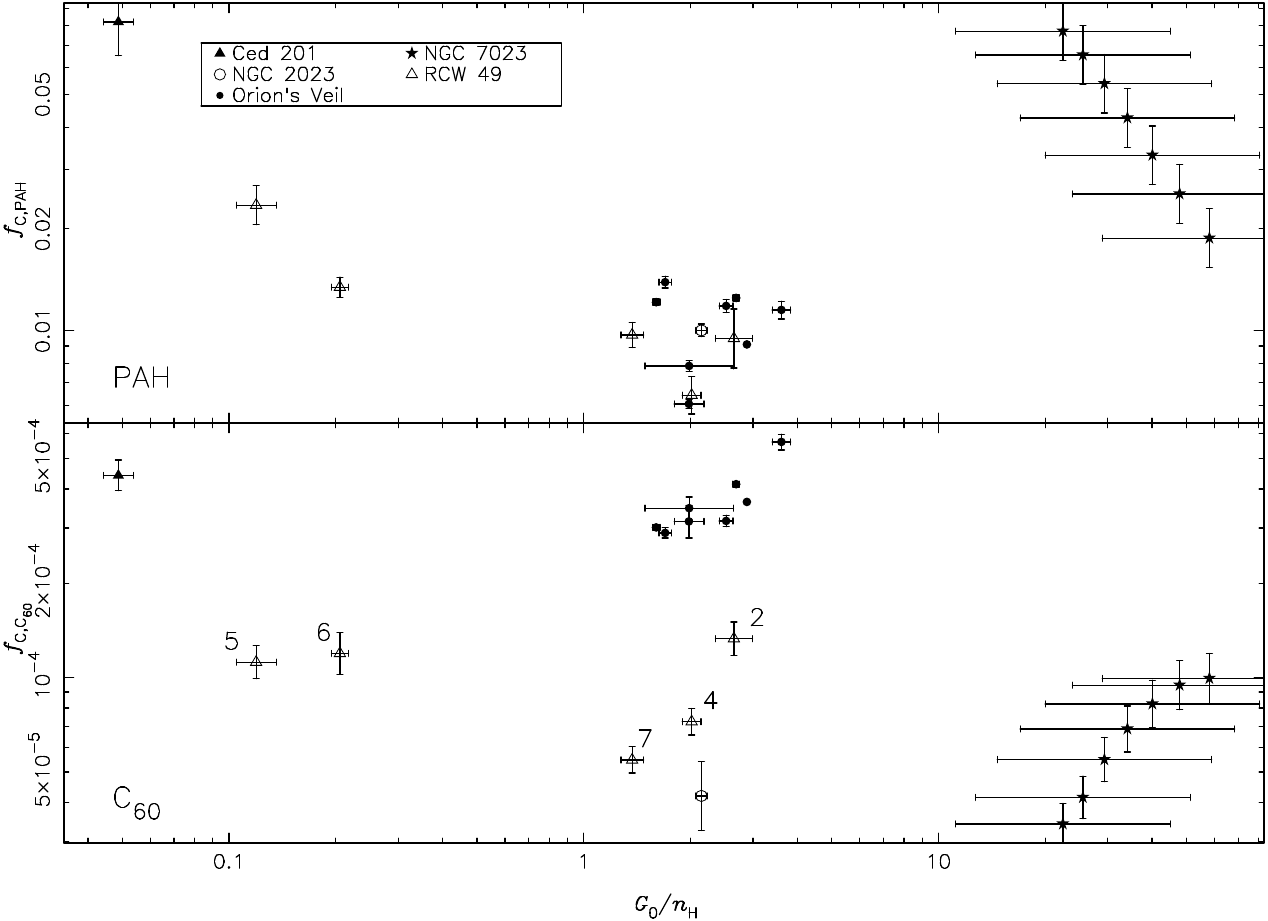}
  \caption{Variations of the carbon fraction locked in PAHs (top) and
    C$_{60}$ (bottom) with respect to $G_0/n_\mathrm{H}$. The
    different symbols indicate the different PDRs included in this
    work for which C$_{60}$ is detected. For RCW~49, the numbers
    correspond to the individual regions considered.}
  \label{fig:abund}
\end{figure*}

\begin{figure*}[t!]
  \centering
  \includegraphics[width=\textwidth,height=!]{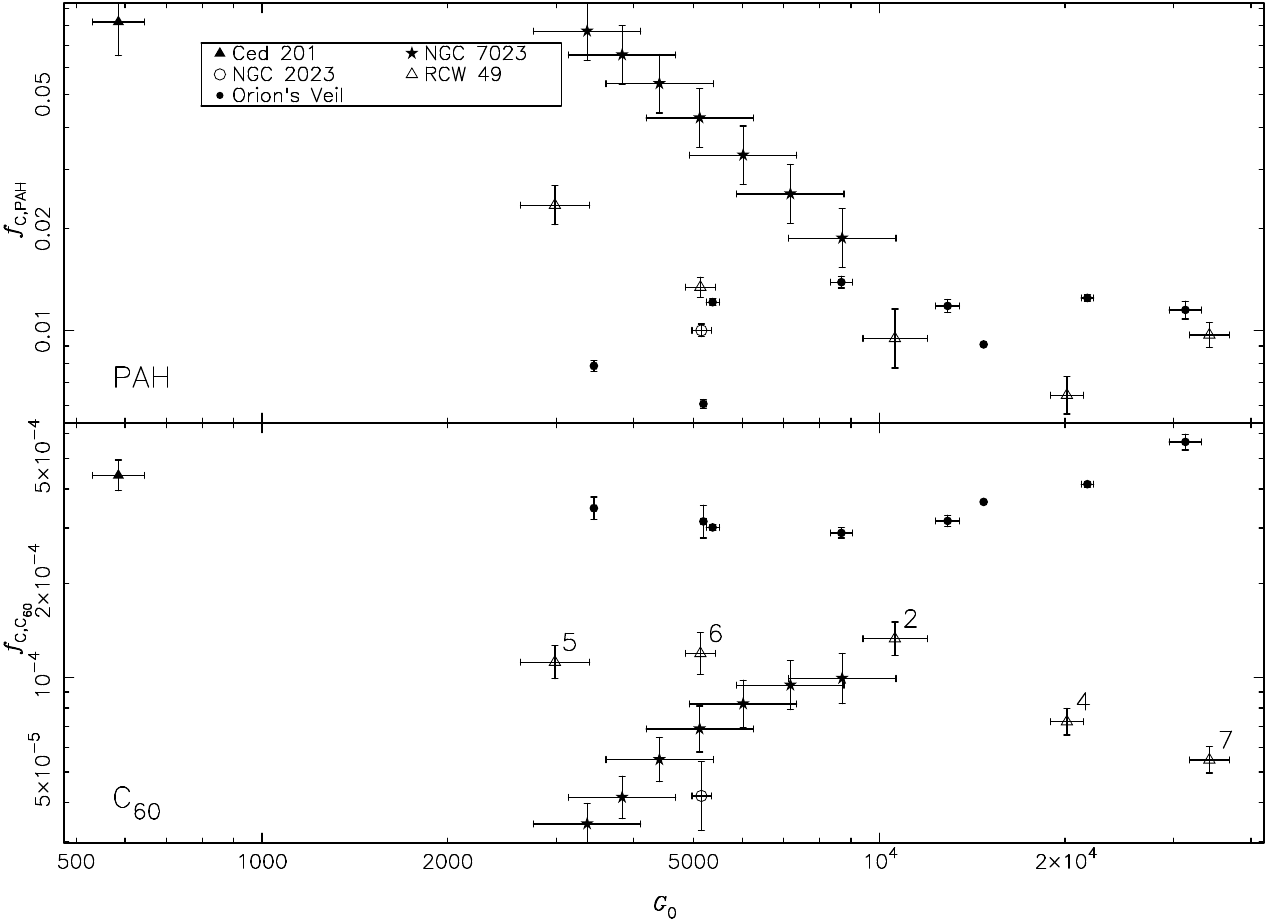}
  \caption{Variations of the carbon fraction locked in PAHs (top) and
    C$_{60}$ (bottom) with respect to $G_0$. The different symbols
    indicate the different PDRs included in this work for which
    C$_{60}$ is detected. For RCW~49, the numbers correspond to the
    individual regions considered.}
  \label{fig:abund_G0}
\end{figure*}

In this section we present results of the variations in abundance of
both PAHs and C$_{60}$. We calculate the abundances from the ratio of
intensity of PAHs and C$_{60}$ with respect to $G_0$, which in turn is
related to the FIR intensity of large dust grains. Assuming that
PAHs, C$_{60}$ and dust compete for the same photons, we can
write \citet[Sec.~6.7]{tie05},
\begin{equation}
  \label{eq:abund}
  f_\mathrm{C} = 0.23\left(\frac{7\times
      10^{-18}~\mathrm{cm}^2}{\sigma_\mathrm{FUV}}\right)\,
  \frac{f_\mathrm{IR}}{1-f_\mathrm{IR}}, 
\end{equation}
where $f_\mathrm{IR}$ is the ratio of the PAHs or C$_{60}$ total
intensities to the total IR intensity \citep[][given by
eq.~(\ref{eq:g0}) for the FIR and adding the contributions from PAHs
and C$_{60}$, following]{ber12}, $\sigma_\mathrm{FUV}$ is the FUV
absorption cross section per carbon atom of the considered
species. Finally, $f_\mathrm{C}$ is the fraction of elemental carbon
locked in the species considered, which is in turn related to the
total abundance.

In this calculation, we consider the total PAH intensity as the
addition of all the PAH bands below 15~$\mu$m. For NGC~2023 and
Ced~201 this includes features starting from $\lambda = 5$~$\mu$m. In
the case of RCW~49, NGC~7023 and Orion's Veil our coverage is limited
to the range between 10--15~$\mu$m since the data in these cases have
been done using the SH mode. Considering the contribution that this
range has to the total PAH intensity in NGC2023 and Ced~201 we find
that it is fairly constant fraction at $\sim 0.2$. We will use this
value as a correction factor in order to estimate the total PAH
abundance in RCW~49, NGC~7023 and Orion's Veil. A final correction is
needed also for the SH data arising from the difference in flux
measurements with respect of the LL data. Using the flux of the
11.2~$\mu$m feature from NGC~7023 in both LL and SH modes we find this
additional factor to be $\sim 3$. Given the mismatch between the
resolution of our $G_0$ maps and the C$_{60}$ and PAH intensity maps,
we re-project and convolve the intensity maps to the PACS 160~$\mu$m
resolution, which is the same as the resolution of our $G_0$ maps.

In order to calculate the total intensity of C$_{60}$ and PAHs at the
position of the corresponding PDRs, we measure $I_{18.9}$,
$I_\mathrm{PAH}$ and $I_\mathrm{FIR}$ at the position were the
combined emission from the available PAH bands peaks. We need to
include the contributions to $I_\mathrm{C_{60}}$ from the bands at
17.4~$\mu$m (which also has contributions from PAHs), 8.5 and
7.0~$\mu$m which we have not detected separately. Considering a first
order approximation, we use the value of 0.6 given by \citet{pee12}
for the ratio of the 17.4 and the 18.9~$\mu$m band intensities, and
0.4 for the ratio of both, the 7.0 and 8.5~$\mu$m with respect to
18.9~$\mu$m \citep{ber12}. The exact value of these ratios, however,
is not precisely determined \citep{bern12}, but even when considering
other calculated values for the intrinsic rates, we find differences
of at most a factor of two.

Using the carbon fractions derived in the previous section, we compare
the results for our sources with other values found in the
literature. In NGC~7023 the observation of growth in C$_{60}$ ranges
from $10^{-5}$ to $10^{-4}$ \citep{ber12}. A study of C-rich, H-poor
PNe by \citet{gar11} shows a range in the carbon fraction in C$_{60}$
from $3\times 10^{-5}$ up to $3\times 10^{-3}$, with most objects
falling in the range of $\sim 10^{-4}$. All these previous estimations
include the contributions to $I_\mathrm{C_{60}}$ from the bands at
17.4~$\mu$m (which also has contributions from PAHs), 8.5 and
7.0~$\mu$m which we have not detected separately. The value of all the
regions considered in this study range fall in the $f_\mathrm{C} =
3\times 10^{-5}-6\times 10^{-4}$, which is in good agreement with the
values found in the previously mentioned studies.

The results of $f_\mathrm{C}$ for both PAHs and C$_{60}$ in the
different PDRs with respect to $G_0/n_\mathrm{H}$ are presented in
Fig.~\ref{fig:abund}. For PAHs, we observe a decrease with increasing
$G_0/n_\mathrm{H}$, for $G_0/n_\mathrm{H} < 1$, from $f_\mathrm{C}
\sim 0.08$ to $f_\mathrm{C} \sim 0.01$. For $G_0/n_\mathrm{H} \sim
1-4$ there is substantial scatter, even within single sources. In
contrast, the case of NGC~7023 shows a clear trend, with higher
abundance than in other regions in the range $10 < G_0/n_\mathrm{H} <
100$. The behavior observed in NGC~7023 appears to be the same as that
observed for the other regions in the range of $G_0/n_\mathrm{H} < 1$.

For C$_{60}$ we do not observe a general trend that can be applied to
all the regions. However, within individual objects we observe that
for $G_0/n_\mathrm{H} > 1$ there is an increase in the abundance for
RCW~49, NGC~7023 and Orion's Veil. In the case of RCW~49 the increase
in abundance does not apply for the full range. For $G_0/n_\mathrm{H}
< 1$ we find that the abundance of C$_{60}$ within RCW~49 appears to
remain constant. Furthermore, even for $G_0/n_\mathrm{H} > 1$ the
trend is not clear, particularly considering that it includes regions
2 and 4, which have poorly determined hydrogen densities. In NGC~7023
and Orion's Veil the increase holds for the full probed range,
although for none of these regions we have points with
$G_0/n_\mathrm{H} < 1$. We cannot conclude if this behavior applies as
well for NGC~2023 or Ced~201 since we have single points for these
PDRs.

Fig.~\ref{fig:abund_G0} shows again $f_\mathrm{C}$ variations of both
species, in this case with respect to $G_0$. For the PAHs we observe a
decrease in abundance in NGC~7023 and, to a lesser extent, in
RCW~49. The abundance in Orion's Veil is consistent with a fairly
constant value or a slight increase. The case of C$_{60}$ abundance
again shows trends within individual regions. In NGC~7023 and Orion's
Veil there is an increase in C$_{60}$ abundance along the probed
range. In Orion's Veil this increase is observed at $G_0 > 10^4$,
while for NGC~7023 the increase seems to halt when $G_0$ approaches
$10^4$. On the other hand, RCW~49 shows a decrease in C$_{60}$
abundance, which also begins at $G_0 \sim 10^4$.

\section{DISCUSSION}\label{sec:disc}

Perusing Figs.~\ref{fig:abund} and \ref{fig:abund_G0}, we conclude
that within NGC~7023, there is a clear trend for a decrease in PAH
abundance and an increase in C$_{60}$ abundance with increasing
$G_0/n_\mathrm{H}$. The other regions show a more mixed behavior. For
RCW~49, for example, the PAH abundance decreases with increasing
$G_0/n_\mathrm{H}$ or $G_0$, but the trend in the C$_{60}$ abundance
is less than clear, while for Orion's Veil the opposite is true. We
attribute the clearness of the trend in NGC~7023 to the fact that the
PDR is seen edge-on, while for other regions the morphological and
geometrical characteristics are more uncertain and might play a role
in the observed trends or lack thereof. Combining all sources together
there is no obvious trend whatsoever in either species or with either
variable. From this, we conclude that besides the physical conditions
-- $G_0$ and $n_\mathrm{H}$ -- there must be other parameter(s)
influencing the evolution of the PAHs and fullerenes.

The clear trends in the PAH and fullerene abundances with $G_0$ in
NGC~7023 have been interpreted in a simple model, describing the
chemical evolution of PAHs under the influence of the stellar UV
photons through H-loss to graphene sheets resulting ultimately, on the
one hand, into small hydrocarbons due to C$_2$ fragmentation and, on
the other hand to fullerenes through isomerization \citep{ber12}. We
note that time is an additional factor entering such chemical
models. Specifically, H-loss is expected to be described by a balance
between collisional hydrogenation and UV-driven dehydrogenation and
thus regulated by $G_0$ and $n_\mathrm{H}$. However, C$_2$-loss is
likely to be a time-dependent process controlled by $G_0$. Hence the
absence of a general trend in the PAH and fullerene abundances across
many sources can be seen as a consequence of this
time-dependence. Moreover, PAH abundance appears to be related to the
age of the PDRs. We observe the highest PAH abundance in Ced~201 and
NGC 7023, the youngest sources in our sample, with respective ages of
1500 and $10^5$~yr \citep{ale08}. For the remaining PDRs we find
similar, overlapping age estimates: 1--7~Myr for NGC~2023
\citep{lop13}, 2--3~Myr for RCW~49 \citep{pia98} and 1--3~Myr for
Orion's Veil \citep{fla03}. For these PDRs we also find lower
abundance of PAHs than in Ced~201 or NGC~7023, which is consistent
with a time-dependent destruction of PAHs.

\begin{figure}[t!]
  \centering
  \includegraphics[width=\columnwidth,height=!]{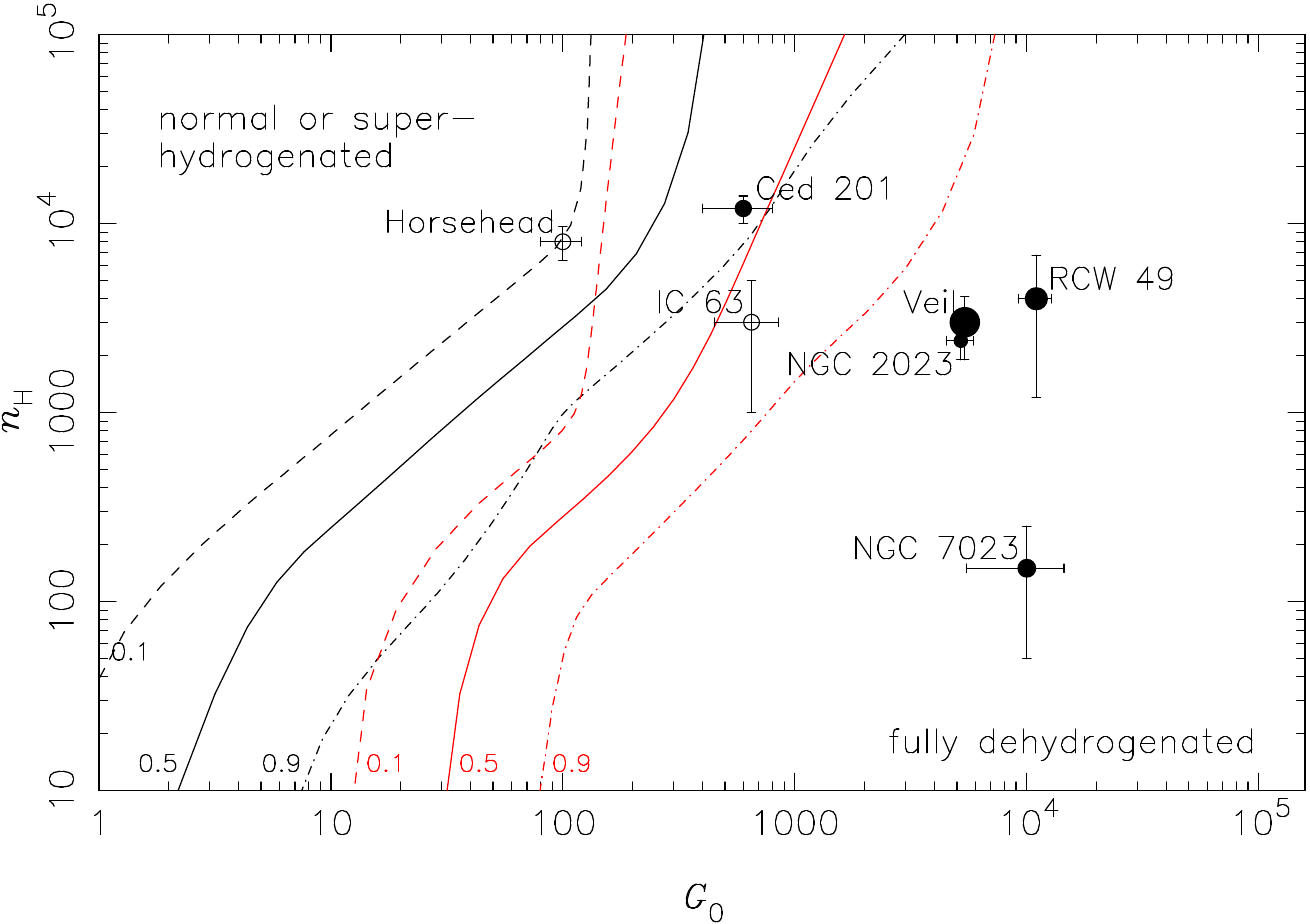}
  \caption{Hydrogenation state of circumcoronene (C$_{54}$H$_{18}$;
    black) and circumovalene (C$_{66}$H$_{20}$; red) as function of
    $G_0$ and the atomic hydrogen density. The labels of the lines
    indicate the fraction of the respective molecules that is fully
    dehydrogenated. The full circles correspond to the PDRs where we
    detect C$_{60}$, the symbol size is proportional to the
    C$_{60}$-to-PAH ratio. The sources for which we can only derive
    upper limits are marked by open circles. The error bars mark the
    ranges for the physical conditions. \citep[Figure adapted from
    ][]{mon13}.}
  \label{fig:dehyd}
\end{figure}

Recently, the first step in the processing from PAHs to fullerenes has
been modeled in detail by \citet{mon13}. The results of this model are
shown in Fig.~\ref{fig:dehyd}. On the left-hand-side of this figure,
PAHs are fully hydrogenated while the right-hand-side corresponds to
completely dehydrogenated PAHs -- i.e. graphene flakes. More specific,
the lines in this figure represent the loci at which the
hydrogenation-balance leads to a constant H-fraction on a PAH of a
given size. So, the further to the right, the more of the initial PAH
size distribution will be destroyed but as long as the PAH size
distribution extends to $\sim 100$ C-atoms, some fraction of the PAHs
will remain hydrogenated. The different sources in our sample are
indicated by symbols in this figure. We note that the sources without
detectable C$_{60}$ are to the left of the ``stability-line'' for PAHs
in excess of 60 C-atoms, leading some credence to a photochemical
transformation model of PAHs into fullerene in the ISM. However, the
C$_{60}$/PAH abundance ratio does not correlate well with the position
into the zone where PAHs of the relevant sizes are expected to be
fully dehydrogenated, as can be seen in Fig.~\ref{fig:dehyd}. This may
merely indicate that time is of the essence or that there are real
fluctuations in the PAH family from region to region. Ced~201 forms an
interesting exception to this as it has C$_{60}$ despite being in the
``stable-zone'' for PAHs with more than 60 C-atoms. This is however a
very peculiar source as it is a chance encounter of the illuminating
star with a cloud. This star-cloud interaction seems to have led to a
shock wave \citep{ces00} located in the position where the C$_{60}$
abundance is particularly high. Hence, it is tempting to speculate
that in this particular source C$_{60}$ is formed by shock processing
of PAHs and/or dust grains. The inferred shock velocity is quite low
\citep[$\sim 10$~km~s$^{-1}$][]{wit87} and such low velocity shocks
are not expected to lead to much PAH destruction \citep{mic10}. On the
other hand, the observed C$_{60}$/PAH abundance is only 0.005 and
hence little processing is required.

Finally, small HAC grains have also been considered as photochemical
starting points for C$_{60}$ formation in the ISM. The general
processes involved are the same: the first step requires
dehydrogenation of the HAC particles followed by a shrinking of the
resulting cage \citep{mic12}. As for PAHs, the hydrogenation of HAC is
set by a balance between reactions with H and photochemical H-loss
\citep{men01,men02,men06} and therefore controlled by
$G_0/n_\mathrm{H}$ \citep{chi13}. The main difference with the PAH
model resides in the timescale for C-loss. Based upon molecular
dynamics calculations \citep{zhe07}, this C-loss timescale is
calculated to equal the IR-cooling timescale at 100--200~K for HACs
and such a low temperature is readily attained by HAC grains
containing $\simeq 500$~C-atoms (or less). However, the adopted
C-binding energies are only 0.36 eV \citep{mic12} and this is much
less than expected for carbon cages. As emphasized in the original
molecular dynamics study, these results may not be statistically
significant \citep[S.\ Irle, private communication;][]{zhe07}. Further
laboratory studies will have to settle this issue.

\section{CONCLUSIONS}\label{sec:concl}

We present a survey of C$_{60}$ in PDRs. While in NGC~2023 the
presence of fullerenes had already been established
\citep{sel10,pee12}, the detection on RCW~49 and Ced~201 represent new
sources where C$_{60}$ is confirmed to be present. We also quantified
the abundances as fraction of elemental carbon and found values
consistent with other studies of C$_{60}$ in PDRs, with values ranging
from $\sim 3\times 10^{-5}$ to $\sim 6\times 10^{-4}$. Furthermore,
the values we derive for $T_d$ from FIR observations indicate that
these regions have temperatures ranging from 20 to 40~K, which is too
low for C$_{60}$ to be emitting in grains, and supports the idea of a
gas phase species or small sized clusters undergoing stochastic
heating \citep{sel10}. The well-known strong IR bands at 6.2, 7.7,
8.6, 11.2, 12.7 and 16.4~$\mu$m, show a very similar spatial behavior,
with minor variations in all sources. In contrast, the spatial
distribution of the 18.9~$\mu$m band is very different and we conclude
that this band has a different carrier \citep[i.e.,
C$_{60}$][]{cam10,sel10} than the other bands (i.e., PAHs).

While some of the sources appear to have trends, we find no relation
between C$_{60}$ or PAH abundance and either $G_0/n_\mathrm{H}$ or
$G_0$ when considering all PDRs together. We consider age as a factor
explaining the lack of a general trend. Comparing our observational
results to the model predictions of \citet{mon13} for PAH
dehydrogenation, we find that regions where C$_{60}$ is detected have
physical conditions consistent with full dehydrogenation of PAHs of at
least 60 C-atoms, with the exception of Ced~201. The conditions of
regions where only upper limits for C$_{60}$ abundance could be
derived, support only partial dehydrogenation of PAHs of the relevant
size. These results support models where the dehydrogenation of
carbonaceous species is the first step towards C$_{60}$ formation
\citep{ber12,mic12}.

More observations aimed at measuring variations of C$_{60}$ abundance
with respect to both PAHs and HAC, as well as better determinations of
$n_\mathrm{H}$ are needed to confirm the reality of these trends and
the more likely parent species. Better constraints on the age
estimates and the study of additional PDRs with significant age
differences are needed to test our hypothesis about the effect of age
in generating the seemingly independent trends for each PDR.

\begin{acknowledgements}

  Studies of interstellar chemistry at Leiden Observatory are
  supported through advanced-ERC grant 246976 from the European
  Research Council, through a grant by the Dutch Science Agency, NWO,
  as part of the Dutch Astrochemistry Network, and through the Spinoza
  premie from the Dutch Science Agency, NWO. P.C. acknowledges support
  from a Huygens fellowship at Leiden University.

  We would like to thank B.\ Ochsendorf and H.\ Andrews for
  facilitating us their reduced spectra of B~33 and IC~63
  respectively. We would also like to thank J.\ Montillaud for the
  data on dehydrogenation state of PAHs. We also acknowledge the
  anonymous referee, whose comments helped us in improving this paper.

  This work is based in part on observations made with the {\it
    Spitzer Space Telescope}, which is operated by the Jet Propulsion
  Laboratory, California Institute of Technology under a contract with
  NASA.

  The {\it Herschel} spacecraft was designed, built, tested, and
  launched under a contract to ESA managed by the Herschel/Planck
  Project team by an industrial consortium under the overall
  responsibility of the prime contractor Thales Alenia Space (Cannes),
  and including Astrium (Friedrichshafen) responsible for the payload
  module and for system testing at spacecraft level, Thales Alenia
  Space (Turin) responsible for the service module, and Astrium
  (Toulouse) responsible for the telescope, with in excess of a
  hundred subcontractors.

  PACS has been developed by a consortium of institutes led by MPE
  (Germany) and including UVIE (Austria); KU Leuven, CSL, IMEC
  (Belgium); CEA, LAM (France); MPIA (Germany); INAF-IFSI/OAA/OAP/OAT,
  LENS, SISSA (Italy); IAC (Spain). This development has been
  supported by the funding agencies BMVIT (Austria), ESA-PRODEX
  (Belgium), CEA/CNES (France), DLR (Germany), ASI/INAF (Italy), and
  CICYT/MCYT (Spain).

\end{acknowledgements}


\begin{thebibliography}

\bibitem[Abergel et al.(2003)]{abe03} Abergel, A., Teyssier, D.,
  Bernard, J.P., et al.\ 2003, A\&A, 410, 577

\bibitem[Alecian et al.(2008)]{ale08}Alecian, E., Catala, C., Wade,
  G.A., et al.\ 2008, MNRAS, 385, 391

\bibitem[Bernard-Salas et al.(2012)]{bern12}Bernard-Salas, J., Cami,
  J., Peeters, E., et al.\ 2012, ApJ, 757, 41

\bibitem[Bern\'e et al.(2013)]{ber13}Bern\'e, O., Mulas, G.\ \&
  Joblin, C.\ 2013, A\&A, 550, L4

\bibitem[Bern\'e \& Tielens(2012)]{ber12}Bern\'e, O.\ \& Tielens,
  A.G.G.M.\ 2012, PNAS, 109, 401

\bibitem[Boersma et al.(2012)]{boe12}Boersma, C., Rubin, R.H.\ \&
  Allamandola, L.J.\ 2012, ApJ, 753, 168

\bibitem[Brand \& Blitz(1993)]{bra93}Brand, J.\ \& Blitz, L.\ 1993,
  A\&A, 275, 67

\bibitem[Brown et al.(1994)]{bro94}Brown, A.G.A., de Geus, E.J.\ \&
  de Zeeuw, P.T.\ 1994, A\&A, 289, 101

\bibitem[Cami et al.(2010)]{cam10}Cami, J., Bernard-Salas, J.,
  Peeters, E.\ \& Malek, S.E.\ 2010, Sci, 329, 1180

\bibitem[Casey(1991)]{cas91}Casey, S.C.\ 1991, ApJ, 371, 183

\bibitem[Carraro et al.(2013)]{car13}Carraro, G., Turner, D.,
  Majaess, D.\ \& Baume, G.\ 2013, A\&A, 555, A50

\bibitem[Cesarsky et al.(2000)]{ces00}Cesarsky, D., Lequeux, J.,
  Ryter, C.\ \& G\'erin, M.\ 2000, A\&A, 354, L87

\bibitem[Chase et al.(1992)]{cha92}Chase, B., Herron, N.\ \& Holler,
  E.\ 1992, JPhCh, 96, 4262

\bibitem[Cherchneff et al.(2000)]{che00}Cherchneff, I., Le Teuff,
  Y.H., Williams, P.M.\ \& Tielens, A.G.G.M.\ 2000, A\&A, 357, 572

\bibitem[Chiar et al.(2013)]{chi13}Chiar, J.E., Tielens,
  A.G.G.M., Adamson, A.J.\ \& Ricca, A.\ 2013, ApJ, 770, 78

\bibitem[Choi et al.(2000)]{cho00}Choi, C.H., Kertesz, M.\ \& Mihaly,
  L.\ 2000, JPCA, 104, 102

\bibitem[Churchwell et al.(2004)]{chu04}Churchwell, E., Whitney,
  B.A., Babler, B.L., et al.\ 2004, ApJS, 154, 322

\bibitem[Draine(2003)]{dra03}Draine, B.T.\ 2003, ARA\&A, 41, 241

\bibitem[Dupac et al.(2003)]{dup03}Dupac, X., del Burgo, C., Bernard,
  J.-P., et al.\ 2003, A\&A, 404, L11

\bibitem[Fabian(1996)]{fab96}Fabian, J.\ 1996, PhRvB, 53,
  13864

\bibitem[Flaccomio et al.(2003)]{fla03}Flaccomio, E., Damiani, F.,
  Micela, G., et al.\ 2003, ApJ, 582, 398

\bibitem[Foing \& Ehrenfreund(1994)]{foi94}Foing, B.H.\ \&
  Ehrenfreund, P.\ 1994, Natur, 369, 296

\bibitem[Fuente et al.(1995)]{fue95}Fuente, A., Mart\'in-Pintado, J.\
  \& Gaume, R.\ 1995, ApJL, 442, L33

\bibitem[Garc\'ia-Hern\'andez et
  al.(2011)]{gar11}Garc\'ia-Hern\'andez, D.A., Iglesias-Groth, S.,
  Acosta-Pulido, J.A., et al.\ 2011, ApJL, 737, L30

\bibitem[Gielen et al.(2011)]{gie11}Gielen, C., Cami, J., Bouwman,
  J., Peeters, E.\ \& Min, M.\ 2011, A\&A, 536, 54

\bibitem[Habing(1968)]{hab68}Habing, H.J.\ 1968, BAN, 19, 421

\bibitem[Harvey et al.(1980)]{har80}Harvey, P.M., Thronson, H.A.,
  Jr.\ \& Gatley, I. 1980, ApJ, 235, 894

\bibitem[H\o g et al.(2000)]{hog00}H\o g, E., Fabricius, C.,
  Jr., Makarov, V.V., et al.\ 2000, A\&A, 355, L27

\bibitem[Houck et al.(2004)]{hou04}Houck, J.R., Roellig, T.L., van
  Cleve, J., et al.\ 2004, ApJS, 154, 18

\bibitem[Iglesias-Groth et al.(2011)]{igl11}Iglesias-Groth, S.,
  Cataldo, F.\ \& Manchado, A.\ 2011, MNRAS, 413, 213

\bibitem[Ijima(1991)]{iji91}Ijima, S.\ 1991, Natur, 354, 56

\bibitem[Jansen et al.(1994)]{jan94}Jansen, R.A., van Dishoeck, E.F.\
  \& Black, J.H.\ 1994, A\&A, 282, 605

\bibitem[Kemper et al.(1999)]{kem99}Kemper, C., Spaans, M., Jansen,
  D.J., et al.\ 1999, ApJ, 515, 649

\bibitem[Knapp et al.(1975)]{kna75}Knapp, S.L., Brown, R.L.\ \&
  Kuiper, T.B.H.\ 1975, ApJ, 196, 167

\bibitem[Kr\"atschmer et al.(1990)]{kra90}Kr\"atschmer, W., Lamb,
  L.D., Fostiropoulos, K.\ \& Huffman, D.R.\ 1990, Natur, 347, 354

\bibitem[Kroto et al.(1985)]{kro85}Kroto, H.W., Heath, J.R., O'Brien,
  S.C., Curl, R.F.\ \& Smalley, R.E.\ 1985, Natur, 318, 162

\bibitem[Lada et al.(1991)]{lad91}Lada, E.A., DePoy, D.L., Evans,
  N.J., II \& Gatley, I.\ 1991, ApJ, 371, 171

\bibitem[L\'opez-Garc\'ia et al.(2013)]{lop13}L\'opez-Garc\'ia, M.A.,
  L\'opez-Santiago, J., Albacete-Colombo, J.F., P\'erez-Gonz\'alez,
  P.G.\ \& de Castro, E.\ 2013, MNRAS, 429, 775

\bibitem[Maier(1994)]{mai94}Maier, J.P.\ 1994, Natur, 370, 423

\bibitem[Men\'endez \& Page(2000)]{men00}Men\'endez, J.\ \& Page,
  J.B.\ 2000, in Light Scattering in Solids VIII: Fullerenes,
  Semiconductor Surfaces, Coherent Phonons, eds.\ Cardona, M.\ \&
  G\"untherodt (Berlin: Springer), 27

\bibitem[Mennella(2006)]{men06}Mennella, V.\ 2006, ApJL, 647, L49

\bibitem[Mennella et al.(2002)]{men02}Mennella, V., Brucato, J.R.,
  Colangeli, L.\ \& Palumbo, P.\ 2002, ApJ, 569, 531

\bibitem[Mennella et al.(2001)]{men01}Mennella, V., Mu\~noz Caro, G.M.,
  Ruiterkamp, R., et al.\ 2001, A\&A, 367, 355

\bibitem[Micelotta et al.(2012)]{mic12}Micelotta, E.R., Jones, A.P.,
  Cami, J., et al.\ 2012, ApJ, 761, 35

\bibitem[Micelotta et al.(2010)]{mic10}Micelotta, E.R., Jones, A.P.\
  \& Tielens, A.G.G.M.\ 2010, A\&A, 510, A36

\bibitem[Moffat et al.(1991)]{mof91}Moffat, A.F.J., Shara, M.M.\ \&
  Potter, M.\ 1991, AJ, 102, 642

\bibitem[Montillaud et al.(2013)]{mon13}Montillaud, J., Joblin, C.\
  \& Toublanc, D.\ 2013, A\&A, 552, 15

\bibitem[Mookerjea et al.(2009)]{moo09}Mookerjea, B., Sandell, G.,
  Jarrett, T.H.\ \& McMullin, J.P.\ 2009, A\&A, 507, 1485

\bibitem[Ohama et al.(2010)]{oha10}Ohama, A., Dawson, J.R., Furukawa,
  N., et al.\ 2010, ApJ, 709, 975

\bibitem[Otsuka et al.(2012)]{ots12}Otsuka, M., Kemper, F., Sargent,
  B., et al.\ 2012, in ASP Conf. Ser. 458, Galactic Archaeology:
  Near-Field Cosmology and the Formation of the Milky Way, ed.\ W.\
  Aoki, M.\ Ishigaki, T.\ Suda, T.\ Tsujimoto \& N.\ Aimoto (San
  Francisco, CA: ASP), 137

\bibitem[Pankonin \& Walmsley(1976)]{pan76}Pankonin, V.\ \& Walmsley,
  C.M.\ 1976, A\&A, 48, 314

\bibitem[Peeters et al.(2012)]{pee12}Peeters, E., Tielens, A.G.G.M.,
  Allamandola, L.J.\ \& Wolfire, M.G.\ 2012, ApJ, 747, 44

\bibitem[Piatti et al.(1998)]{pia98}Piatti, A.E., Bica, E.\ \&
  Claria, J.J.\ 1998, A\&AS, 127, 423

\bibitem[Pilbratt et al.(2010)]{pil10}Pilbratt, G.L., Riedinger,
  J.R., Passvogel, T., et al.\ 2010, A\&A, 518, L1

\bibitem[Pilleri et al.(2012)]{pil12}Pilleri, P., Montillaud, J.,
  Bern\'e, O.\ \& Joblin, C.\ 2012, A\&A, 542, A69

\bibitem[Poglitsch et al.(2010)]{pog10}Poglitsch, A., Waelkens, C.,
  Geis, N., et al.\ 2010, A\&A, 518, L2

\bibitem[Rauw et al.(2004)]{rau04}Rauw, G., De Becker, M., Naz\'e,
  Y., et al.\ 2004, A\&A, 420, L9

\bibitem[Rauw et al.(2007)]{rau07}Rauw, G., Manfroid, J., Gosset, E.\
  et al.\ 2007, A\&A, 463, 981

\bibitem[Roberts et al.(2012)]{rob12}Roberts, K.R.G., Smith, K.T.\ \&
  Sarre, P.J.\ 2012, MNRAS, 421, 3277

\bibitem[Rubin et al.(2011)]{rub11}Rubin, R.H., Simpson, J.P.,
  O'Dell, C.R., et al.\ 2011, MNRAS, 410, 1320

\bibitem[Sellgren et al.(2010)]{sel10}Sellgren, K., Werner, M.W.,
  Ingalls, J.G., et al.\ 2010, ApJL, 722, L54

\bibitem[Sheffer et al.(2011)]{she11}Sheffer, Y., Wolfire, M.G.,
  Hollenbach, D.J., Kaufman, M.J.\ \& Cordier, M.\ 2011, ApJ, 741, 45

\bibitem[Shetty et al.(2009)]{she09}Shetty, R., Kauffmann, J.,
  Schnee, S.\ \& Goodman, A.A.\ 2009, ApJ, 696, 676

\bibitem[Smith et al.(2007a)]{smi07a}Smith, J.D.T., Armus, L., Dale,
  D.A., et al.\ 2007a, PASP, 119, 1133

\bibitem[Smith et al.(2007b)]{smi07b}Smith, J.D.T., Draine, B.T.,
  Dale, D.A., et al.\ 2007b, ApJ, 656, 770

\bibitem[Steiman-Cameron et al.(1997)]{ste97}Steiman-Cameron, T.Y.,
  Haas, M.R., Tielens, A.G.G.M.\ \& Burton, M.G.\ 1997, ApJ, 478, 261

\bibitem[Thi et al.(2009)]{thi09}Thi, W.-F., van Dishoeck, E.F., Bell,
  T., Viti, S.\ \& Black, J.\ 2009, MNRAS, 400, 662

\bibitem[Tielens(2005)]{tie05}Tielens, A.G.G.M.\ 2005, The Physics
  and Chemistry of the Interstellar Medium (1st ed.; Cambridge:
  Cambridge Univ.\ Press)

\bibitem[Tielens(2008)]{tie08}Tielens, A.G.G.M. 2008, ARA\&A, 46, 289

\bibitem[Werner et al.(2004)]{wer04}Werner, M.W., Roellig, T.L., Low,
  F.J., et al.\ 2004, ApJS, 154, 1

\bibitem[Westerlund(1960)]{wes60}Westerlund, B.\ 1960, ArA,
  2, 419

\bibitem[Whiteoak \& Uchida(1997)]{whi97}Whiteoak, J.B.Z.\ \& Uchida,
  K.I.\ 1997, A\&A, 317, 563

\bibitem[Witt et al.(1987)]{wit87}Witt, A.N., Graff, S.M., Bohlin,
  R.C.\ \& Stecher, T.P.\ 1987, ApJ, 321, 912

\bibitem[Young Owl et al.(2002)]{you02}Young Owl, R.C., Meixner,
  M.M., et al.\ 2002, ApJ, 578, 885

\bibitem[Zhang \& Kwok(2011)]{zha11}Zhang, Y.\ \& Kwok, S.\ 2011,
  ApJ, 730, 126

\bibitem[Zheng et al.(2007)]{zhe07}Zheng, G., Wang, Z., Irle, S.\ \&
  Morokuma, K.\ 2007, J.\ Nanosci.\ Nanotechnol., 7, 1662

\end{thebibliography}
\end{document}